\documentclass[aps,prd,twocolumn,showkeys,showpacs]{revtex4-1}
\usepackage{graphicx}
\usepackage{latexsym}
\usepackage{epsfig}
\usepackage{color}
\usepackage[colorlinks, linkcolor=blue, citecolor=blue, urlcolor=blue]{hyperref}
\usepackage{amsmath,amssymb,bm,color}
\usepackage[normalem]{ulem}
\usepackage{subfigure}
\usepackage{verbatim}
\usepackage[utf8]{inputenc}

\def\bea{\begin{eqnarray}}
\def\eea{\end{eqnarray}}
\def\be{\begin{equation}}
\def\ee{\end{equation}}

\usepackage{verbatim}
\usepackage{hyperref}

\newcommand{\de}{\mathrm d}

\newcommand{\g}{$\gamma$}

\newcommand{\Fermi}{{\sl Fermi}}
\newcommand{\twompz}{{2MPZ}}
\newcommand{\twoMPZ}{{2MPZ \,}}
\newcommand{\twomrs}{{2MRS}}
\newcommand{\twoMRS}{{2MRS \hspace{0.05cm}}}

\newcommand{\WIB}{{WIBRALS \hspace{0.05cm}}}

\newcommand{\cp}{{$C_p$}}
\newcommand{\code}[1]{\texttt{#1}}
\newcommand{\std}{{\sl reference}}
\newcommand{\cpfree}{{\sl free} $C_p$}
\newcommand{\apjs}{Astrophys. J. Suppl.}
\newcommand{\mnras}{Mon. Not. Roy. Astron. Soc.}

\usepackage{lineno}
\begin{document}
\title{Characterizing the local gamma-ray Universe via angular cross-correlations}

\author{Simone Ammazzalorso$^{1,2}$}
\author{Nicolao Fornengo$^{1,2}$}
\author{Shunsaku Horiuchi$^{3}$}
\author{Marco Regis$^{1,2}$}

\affiliation{$^1$ Dipartimento di Fisica, Universit\`{a} di Torino, via P. Giuria 1, I--10125 Torino, Italy}
\affiliation{$^2$ Istituto Nazionale di Fisica Nucleare, Sezione di Torino, via P. Giuria 1, I--10125 Torino, Italy}
\affiliation{$^3$ Center for Neutrino Physics, Department of Physics, Virginia Tech, Blacksburg, VA 24061, USA}

\email{ammazzalorso@to.infn.it, fornengo@to.infn.it, horiuchi@vt.edu, regis@to.infn.it}

\begin{abstract}
With a decade of $\gamma$-ray data from the {\it Fermi}-LAT telescope, we can now hope to answer how well we understand the local Universe at $\gamma$-ray frequencies. On the other hand, with $\gamma$-ray data alone it is not possible to directly access the distance of the emission and to point out the origin of unresolved sources.
This obstacle can be overcome by cross-correlating the $\gamma$-ray data with catalogs of objects with well-determined redshifts and positions. In this work, we cross-correlate {\it Fermi}-LAT skymaps with the 2MPZ catalog to study the local $z<0.2$ $\gamma$-ray Universe, where about 10\% of the total unresolved $\gamma$-ray background is produced.
We find the signal to be dominated by AGN emissions, while star forming galaxies provide a subdominant contribution. Possible hints for a particle DM signal are discussed. 
\end{abstract}
\maketitle

\section{Introduction}
\label{sec:intro}

The extragalactic \g-ray background (EGB) is defined as the the \g-ray emission remaining after the subtraction of all Galactic sources from the \g-ray sky. It should be sourced by various classes of extragalactic \g-ray emitters, including the common star-forming galaxies, Active Galactic Nuclei such as blazars, and cascades of high-energy particle propagation (for a recent review, see Ref.~\cite{Fornasa:2015qua}). Exotic sources, such as dark matter annihilation or decay, can also contribute to this signal. In the era of the {\it Fermi}-LAT satellite, much has been revealed about the origins of the EGB. Some $\sim$3,000 extragalactic \g-ray sources, dominantly blazars, have been resolved \cite{Acero:2015hja}, which explain up to half of the EGB \cite{TheFermi-LAT:2015ykq}, and the number will almost double with the upcoming FL8Y point source catalog. Removing these extragalactic point sources from the EGB leaves a residual, the so-called unresolved (or isotropic) \g-ray background (UGRB) \cite{Ackermann:2014usa}, whose origins remain debated and is the focus of this analysis. 

The large numbers of EGB point sources detected have enabled increasingly sophisticated predictions for their contributions to the UGRB \cite{Ackermann:2011bg,Ajello:2011zi,Ajello:2013lka,DiMauro:2013xta,DiMauro:2013zfa}. Often, these utilize extrapolations of multi-wavelength observations to predict the source behaviors in the faint unresolved end. In parallel, a number of new and complementary techniques have been developed to study the UGRB in a more direct way. These uniquely exploit the sub-threshold information in the spatial distribution of \g-ray photons, and include the techniques of anisotropy \cite{Ando:2005xg,Ackermann:2012uf,Harding:2012gk,Cuoco:2012yf,Fornasa:2012gu,Ando:2013ff,DiMauro:2014wha,Fornasa:2016ohl,Ando:2017alx}, pixel statistics \cite{Dodelson:2009ih,Malyshev:2011zi,Lisanti:2016jub,Zechlin:2016pme,Zechlin:2015wdz,DiMauro:2017ing}, and spatial cross-correlation with tracers of large-scale structure \cite{Xia:2011ax,Camera:2012cj,Ando:2013xwa,Ando:2014aoa,Fornengo:2014cya,Shirasaki:2014noa,Camera:2014rja,Cuoco:2015rfa,Regis:2015zka,Xia:2015wka,Shirasaki:2015nqp,Feng:2016fkl,Troster:2016sgf,Shirasaki:2016kol,Branchini:2016glc,Cuoco:2017bpv,Shirasaki:2018dkz,Hashimoto:2018ztv}. 

Galaxies provide abundant opportunities that allow powerful probes of the local large-scale structure of the Universe. In recent works, {\it Fermi}-LAT data were cross-correlated with a variety of galaxy catalogs, including the SDSS-DR6 quasars, SDSS-DR8 main galaxies, SDSS-DR8 luminous red galaxies, SDSS-DR12 photo-$z$ galaxies, NVSS radiogalaxies, WI$\times$SC galaxies, the 2MASS galaxies, and the 2MPZ subsample of 2MASS galaxies. Positive correlations (at the level of $3$--$5 \sigma$) were detected on angular scales of $\lesssim 1^\circ$ with all but the luminous red galaxy catalogs \cite{Xia:2015wka,Cuoco:2015rfa,Regis:2015zka,Shirasaki:2015nqp}, providing valuable information on the sources behind the UGRB and constraints on dark matter contributions. Tomographic analyses, whereby depth (redshift) information is also utilized in unraveling the sources of the correlation signals \cite{Camera:2014rja}, have been successful in increasing the significance of the measured correlations with some galaxy catalogs to $\sim 10 \sigma$ \cite{Cuoco:2017bpv}. 

In this work, we perform new analyses of the cross-correlation that focus on disentangling the astrophysical and exotic contributions to the UGRB at low redshift. Galaxy observables, e.g., in the $B$- and $K$-bands, provide proxies for the amount of astrophysical activity and dark matter, respectively. Thus they can be used to predict astrophysical background and dark matter signal strengths. In order to capture this information, we exploit the plethora of multi-wavelength data available on galaxies and perform new position cross-correlation analyses using galaxies divided into multiple quadrants of astrophysical and dark matter signal expectations. We work with the 2MASS Photometric Redshift catalog (2MPZ), which consists of cross-matching 2MASS XSC, WISE and SuperCOSMOS all-sky samples, which provide multi-wavelength data in 8 wavelengths (B, R, I, J, H, Ks, W1, W2) for over a million galaxies with distribution peaked at $z=0.07$. Simply put, one expects dark matter to correlate most cleanly with massive yet astrophysically inactive targets, and also in nearby galaxies since competing astrophysical processes peak at higher redshifts.
The fact that dark matter peaks at low-$z$ stems from three competing effects: stronger clustering (namely, higher concentration for dark matter halos) as $z$ decreases, higher average dark matter density 
as $z$ increases (scaling as $(1+z)^3$), and dilution of the observed radiation as $z$ (i.e., distance) increases. The first and the latter effects win over the second (see, e.g., \cite{Camera:2014rja}), and the different redshift distribution of the dark matter signal compared to astrophysical backgrounds is one of the most important features making cross-correlation analyses relevant for constraining the particle dark matter nature.
Generically speaking, the method can probe weakly interacting massive particle (WIMP) dark matter with annihilation cross section around the thermal value (depending on the mass and type of analysis~\cite{Cuoco:2015rfa}), as confirmed also by the work presented here.

This paper is organized as followed. In Section \ref{sec:data}, we describe the \g-ray data and galaxy catalogs used. In Section \ref{sec:measur}, we describe our analysis procedure. In Section \ref{sec:inter} we present our results and provide discussions for the origins of the UGRB. Section \ref{sec:concl} concludes. We provide details of our various validation checks in Appendix \ref{sec:xck}, and treatment of source modeling in Appendices \ref{sec:hod} and \ref{sec:cp}. Throughout, we adopt the Planck cosmology with parameters from Ref.~\cite{Ade:2015xua}.

\section{Data}
\label{sec:data}

The datasets that we employ in our cross-correlation analyses are (i) the first 9-years data release of \g-rays from {\it Fermi}-LAT, for which we consider a broad energy range running from 630 MeV to 1 TeV, and (ii) the \twoMPZ galaxy catalog. The data sets and data selection are described in the next subsections.

\subsection{Fermi-LAT}
\label{sec:fermidata}

\begin{figure*}[t]
  \centering
  \subfigure{\includegraphics[width=0.45\textwidth]{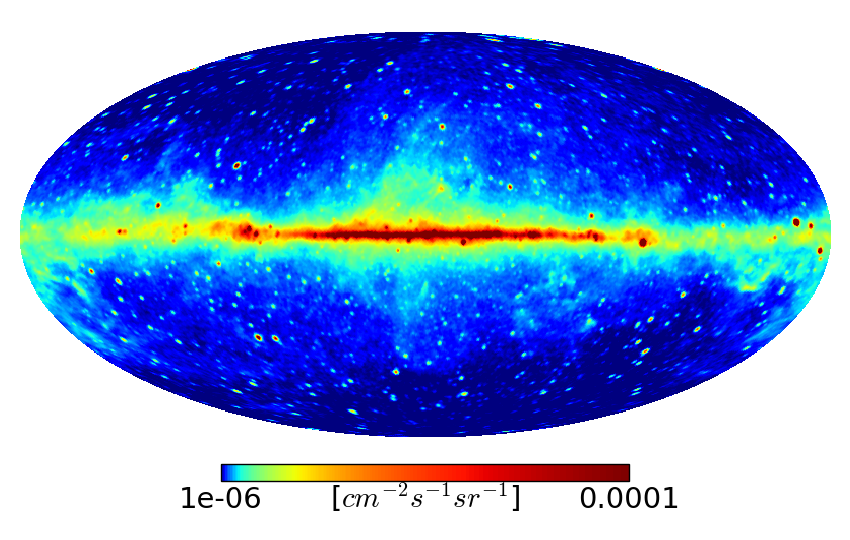}\label{fig:fermimap}}
  \hspace{1cm}
  \subfigure{\includegraphics[width=0.45\textwidth]{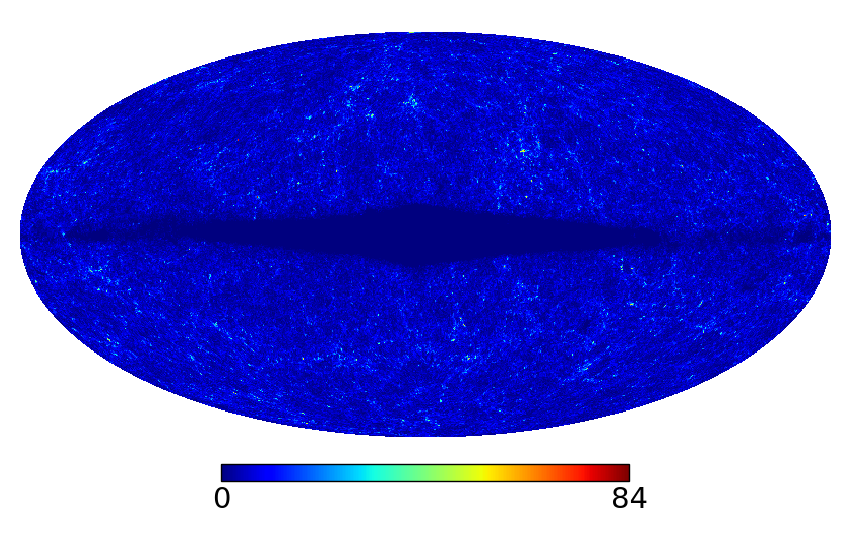}\label{fig:galmap}}
  \caption{Left: All-sky \Fermi-LAT photon intensity map for photon energies above 1 GeV shown in Mollweide projection and smoothed with a Gaussian beam of size $\sigma = 0.4^{\circ}$ for illustration purposes. Right: Galaxy counts map of the full \twoMPZ catalog, in Mollweide projection (the map has been downsized to $N_{\rm side}=128$ for illustration purposes).}
\vspace{0.5cm}
\label{fig:map}
\end{figure*}

{\it Fermi}-LAT is a \g-ray pair-conversion telescope launched in June 2008.
It offers excellent capabilities to investigate the nature of the extra-galactic \g-ray background, covering an energy range between 20 MeV and 1 TeV with remarkable angular resolution ($\sim 0.1^{\circ}$
above 10 GeV) and rejection of charged particles background.

In this work, we use 108 months of data
from August 4th 2008 to July 13th 2017 ({\it Fermi} Mission Elapsed Time: 239557417 s -- 521597050 s).
The photon counts and exposure maps are produced with the LAT Science Tools version v10r0p5\footnote{\href{https://fermi.gsfc.nasa.gov/ssc/data/analysis/software/}{https://fermi.gsfc.nasa.gov/ssc/data/analysis/software/}}. We select the Pass 8\footnote{See \href{http://www.slac.stanford.edu/exp/glast/groups/canda/lat_Performance.htm}{http://www.slac.stanford.edu/exp/glast/groups/canda/ lat\_Performance.htm}, for a definition of the Pass 8 event selections and their features.} ULTRACLEANVETO event class (corresponding to the {\verb"P8R2_ULTRACLEANVETO_V6" instrument response functions (IRFs))}, which is recommended for diffuse emission analysis since it has the lowest cosmic-ray contamination.

We use both back- and front-converting events. For photon energies below 1.2 GeV, where photon statistics is significantly larger than at higher energies but direction reconstruction is worse, we use photons belonging to the event class PSF3, which refers to the best-quality quartile in the reconstructed photon direction (technically, this corresponds to event type 32). At higher energies, for which the direction reconstruction is inherently better but photon statistics declines, we extend the selection to the three best quality quartiles PSF1 + PSF2 + PSF3 (event type 56). This choice allows us to have at the same time a very good angular resolution and a good photon statistics in the whole energy range of our analysis.

The analyses are performed on photon intensity maps, obtained by dividing the count maps by the exposure maps and the pixel area $\Omega_{\rm pix} = 4 \pi / N_{\rm pix} $. We adopt a HEALPix pixelation format with resolution parameter  $N_{\rm side}=1024$, which corresponds to $N_{\rm pix} = 12, 582, 912 $ and a mean spacing  of $\sim 0.06^{\circ}$, similar to the best angular resolution of the gamma-ray data.
Intensity maps are produced in 100 energy bins, evenly spaced in logarithmic scale between 100 MeV and 1 TeV. The resulting intensity maps are then re-binned in larger energy bins for the cross-correlation studies. After various tests, we decided to limit the lowest photon energy to 630 MeV, largely determined by the poorer angular resolution  below this scale. We perform all cross-correlation analyses in 11 energy bins, evenly spaced in logarithmic scale between 630 MeV and 1 TeV, and projected in HEALPix maps with $N_{\rm side}=1024$.

An example of \g-ray map obtained by including all photons above 1 GeV is shown in the Fig.~\ref{fig:map} (left).

\subsubsection{Masking intensity maps}

Although we do not expect a correlation between the Galactic \g-ray diffuse emission and the extra-galactic matter distribution traced by the galaxy catalogs, we nevertheless need to exclude the very bright Galactic emission, especially along the Galactic plane (in order to reduce the noise). We therefore perform a Galactic plane cut by masking galactic latitudes $\vert b \vert<30^{\circ}$ and, in addition, we further subtract the Galactic foreground emission from the data maps.
Resolved point sources are also masked, in order to leave in the intensity maps only the components contributing to the UGRB.

\begin{figure*}[t]
  \centering
  \subfigure{\includegraphics[width=0.45\textwidth]{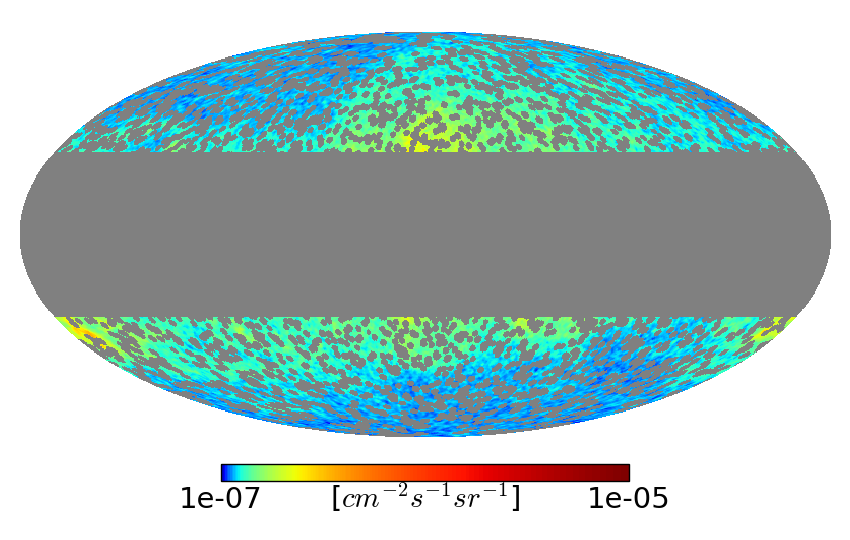}\label{fig:masked}}
  \hspace{1cm}
  \subfigure{\includegraphics[width=0.45\textwidth]{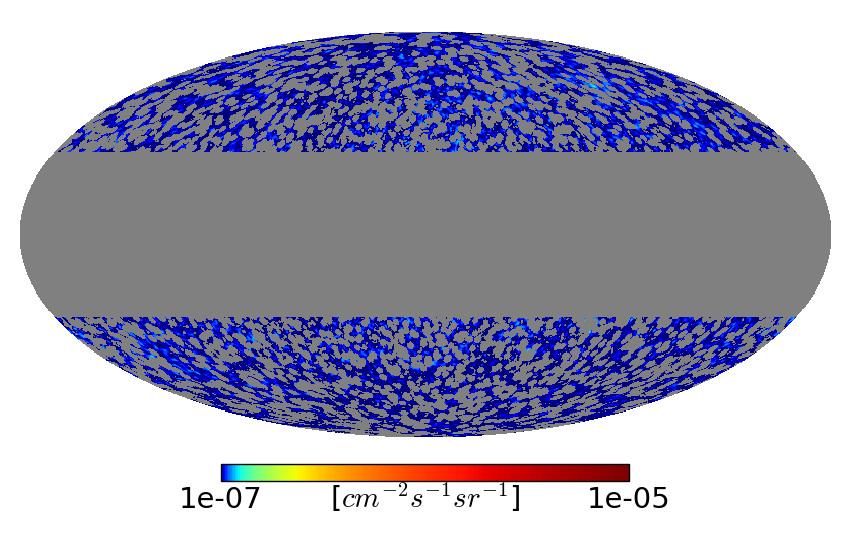}\label{fig:subtracted}}
  \caption{Left: Instance of a masked intensity map: the plots refer to the energy bin $(1.2, 2.3)$ GeV. Gray pixels define the mask, which covers Galactic latitudes $b<30^\circ$ and point sources. The left and right panels show the map without and with Galactic foreground removal.}
  \vspace{0.5cm}
  \label{fig:foresub}
\end{figure*}

The point source masks are based on \Fermi-LAT catalogs of resolved sources. 
We select the sources from the FL8Y catalog: this is a preliminary source list released by the \Fermi-LAT Collaboration which contains almost all the pre-identified sources of the 3FGL catalog augmented by new ones. It includes 5523 sources in the 100 MeV -- 1 TeV energy range \footnote{See: \href{https://fermi.gsfc.nasa.gov/ssc/data/access/lat/fl8y/}{https://fermi.gsfc.nasa.gov/ssc/data/access/lat/fl8y/} for further details.}.
For energies above 10 GeV, we additionally mask sources from the 3FHL \cite{3FHL} catalog, which contains 1556 objects characterized in the 10 GeV -- 2 TeV energy range, in order to account for hard-spectrum sources that might be not contained in the FL8Y catalog. 
The mask is built by taking into account both the angular resolution of the detector in each specific energy bin and the brightness of the source to be masked. Specifically, for each source we mask the pixels inside a circle of radius $R$ around its position defined through the following condition:
\be
F_{\Delta E}^\gamma \, \exp{\left(-\frac{R^2}{2 \theta_{\Delta E}^2}\right)} > \frac{F_{\Delta E,\rm faintest}^\gamma}{5}
\label{eq:radius}
\ee
where $F_{\Delta E}^\gamma$ is integral flux of the source in a given energy bin $\Delta E$, $F_{\Delta E,\rm faintest}^\gamma$ is the flux of the faintest source in the same energy bin, and $\theta_{\Delta E}$ is the 68\% containment angle in that energy bin, as provided by the \Fermi-LAT point-spread-function (PSF) analysis.
The threshold condition based on 1/5 of the flux of the faintest source corresponds approximately to the rms in the specific energy bin (sources are detected with TS $ \geq 25$). It guarantees to properly mask the resolved sources and reduce the chance to have artifacts in the angular power spectrum (APS) due to leakage of the source outside the mask. At the same time, the improvement of the \Fermi-LAT PSF with increasing energy allows to set energy-dependent masks which improve (i.e., become progressively less constraining) as energy grows, which is important since it coincides with where photon statistics become reduced. 
Sources that are marked as ``extended'' in the FL8Y/3FHL catalog, are masked with the ``extension radius'' provided in the \Fermi-LAT catalog.
An example of the mask is shown in Fig.~\ref{fig:foresub} (left) for the energy bin $(1.2, 2.3)$ GeV.

Foreground removal is done by using the Galactic emission model {\tt gll\_iem\_v06.fits} of the {\it Fermi}-LAT Collaboration \footnote{\href{https://fermi.gsfc.nasa.gov/ssc/data/access/lat/BackgroundModels.html}{https://fermi.gsfc.nasa.gov/ssc/data/access/lat/\\ BackgroundModels.html}}. Foreground template maps are produced in the same 100 energy bins, evenly spaced in logarithmic scale between 100 MeV and 1 TeV, and projected in HEALPix maps with $N_{\rm side}=1024$ as introduced for the intensity maps. 
Each template map is assigned a free normalization (and added to a free constant, representing the UGRB and cosmic-ray contamination) and a Poissonian likelihood fit is performed globally on all the masked intensity maps. Through this procedure, we obtained that all the best-fit normalization parameters are of the order of unity, supporting a successful description of the foreground emission. 
The normalized foreground templates are then re-binned into the 11 energy bins used for the cross-correlation analyses, and subtracted from the corresponding intensity maps.
The robustness of foreground removal and choice of foreground model are discussed in Appendix~\ref{sec:appfor}.

The masked intensity map in the energy bin $(1.2 - 2.3)$ GeV, after subtraction of the Galactic foreground emission, is shown in the right panel of Fig.~\ref{fig:foresub}.

\subsection{Galaxy catalogs}
\label{sec:catadata}

For our analysis we employ the \twoMPZ catalog \cite{Bilicki14}, which has been built by cross-matching 2MASS XSC, WISE and SuperCOSMOS all-sky samples. The catalog contains $ \sim 10^6$ galaxies and their photometric redshifts have been reconstructed via an artificial neural network approach. All the 8 magnitudes (B, R, I, J, H, Ks, W1, and W2) measured in SuperCOSMOS, 2MASS and WISE are present. 
In order to perform our measurement we use the mask described in Ref.~\cite{Alonso2015}, which avoids systematics due to Galactic dust contamination or misidentification that derives from high stellar number densities.

Our goal is to decipher the composition of the UGRB at low-z. The different \g-ray emitters considered in this work---dark matter (DM), star forming galaxies (SFG), blazars (BLZ) and misaligned active galactic nuclei (mAGN)---can show different levels of correlations with different subsamples of the \twoMPZ catalog that trace different properties of galaxies. 
In fact, different \g-ray sources can have different redshift behaviors and can be hosted by different types of galaxies. Therefore, in an attempt to enhance the sensitivity of the cross-correlation analysis to the different \g-ray type of sources, in some of our analyses we subdivide the galaxy catalog in several subsamples, as described in the following subsection.

\begin{figure*}[t]
  \centering
  \includegraphics[width=0.9\textwidth]{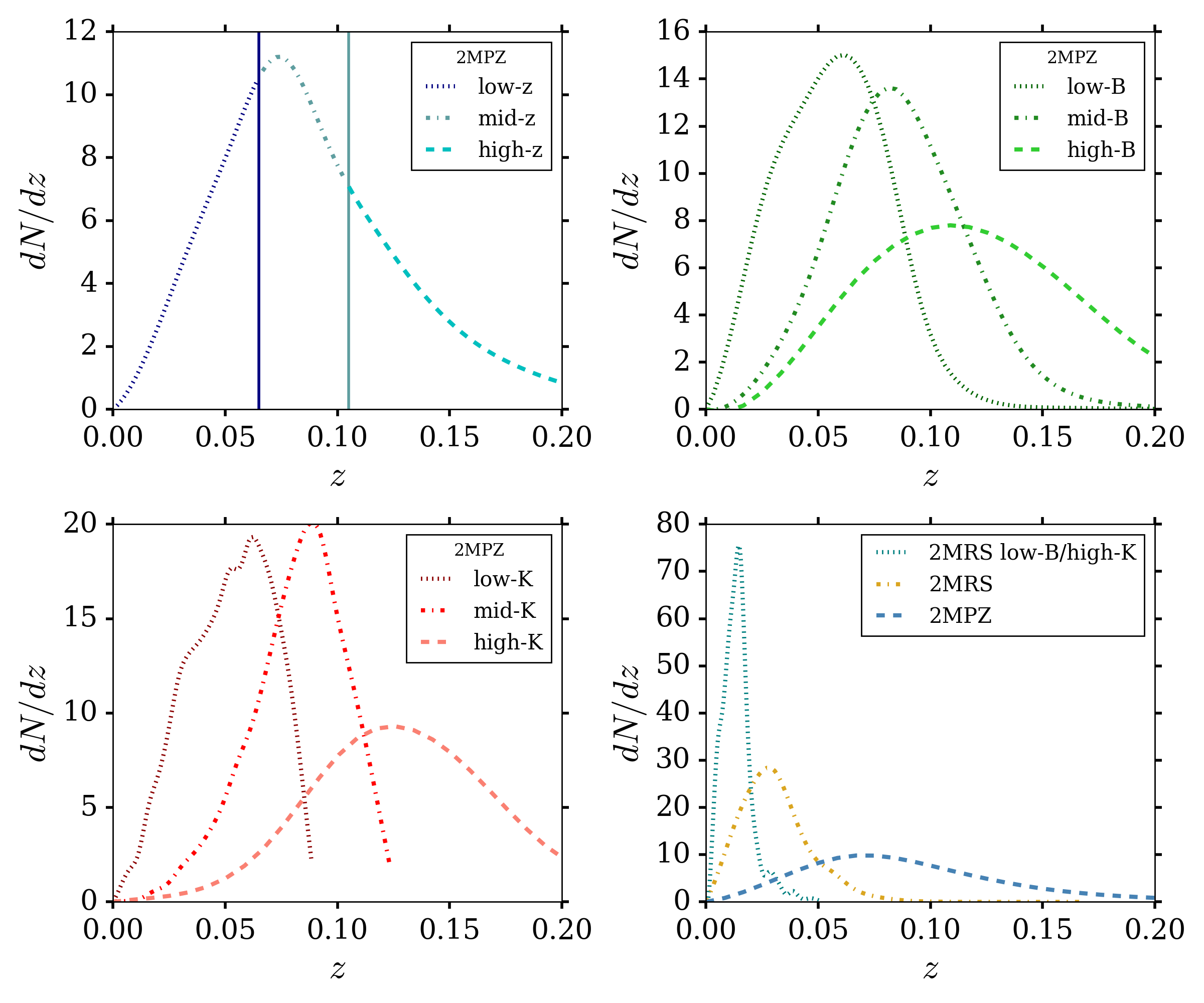}
  \caption{Redshift distributions of the different galaxy subsamples used in the cross-correlation analyses. Each distribution is normalized to the total number of galaxies of its corresponding subsample. Low/Mid/High-B (K) refers to galaxies selected according to their B (or K) luminosity.}
  \label{fig:dNdz}
\end{figure*}

\subsubsection{Galaxy subsets}
\label{subsec:galsub}
The {\bf full 2MPZ} represents our reference catalog, shown in Fig.~\ref{fig:map} (right). 
In addition, we consider the following subsamples:
\begin{itemize}
\item {\bf 2MRS} -- 
The 2MASS Redshift Survey (\twomrs, \cite{2MRS}) contains all the 2MASS sources for which a spectroscopic redshift is available. The catalog counts 50k objects with a mean redshift of $z = 0.03$, thus representing a low-$z$ subsample of \twompz. Since the DM signal is peaked at low redshift, the \twoMPZ has the potential to be more sensitive to the DM \g-ray emission.

\item {\bf Redshift bins} --
We perform a redshift slicing of \twoMPZ subdividing the catalog into three samples ($z<0.07$, $0.07<z<0.11$, $z>0.11$). Each sub-catalog contains approximately one third of the total number of galaxies.

\item {\bf B-luminosity bins} -- 
The B-band luminosity is a reasonable tracer of star formation activity (see, e.g., Ref.~\cite{Bothwell:2009jg}), and thus, would be expected to correlate also with cosmic-ray induced astrophysical \g-ray emission \cite{Ackermann:2012vca}. We thus split the full \twoMPZ catalog into three bins of absolute B-luminosity, with each sub-catalog containing again one third of the total number of galaxies. 

\item {\bf K-luminosity bins} -- 
The K-band luminosity of galaxies are correlated with the stellar mass of the galaxy which can be correlated with the halo mass by, e.g., abundance matching \cite{Behroozi:2010rx}. Therefore, we consider it a tracer of the object mass and we define three sub-catalogs by slicing the full \twoMPZ catalog into three bins of absolute K-luminosity again each one containing one third of the total number of galaxies. 

\item {\bf High K--Low B} -- 
Objects with high-K and low-B luminosities should have high mass and low level of star formation activity. Therefore they can be considered as ideal targets for DM searches, since they might have a reduced correlation with astrophysical \g-ray sources (having the emission driven by star formation activity), whilst an enhanced correlation with \g-ray emission induced by DM (which is related to the mass). In order to perform this investigation, we select 10k objects in the corner of the plane of K vs B absolute luminosity in the \twomrs \ catalog (since the DM signal is peaked at low-$z$). We will report the results about this sample only when focusing on the DM interpretation in Section~\ref{sec:disc}.

\end{itemize}

Fig.~\ref{fig:dNdz} shows the redshift distributions of the full \twoMPZ compared to those of the subsample catalogs. \twomrs\ is the catalog peaking at the lowest redshift. The subsamples of the mid bins in both K and B luminosity have a redshift distribution close to the full 2MPZ, while the low/high bins are peaked at lower/higher $z$. 

In addition to the subsamples listed above, we further define two selections of sources that aim at identifying specifically mAGN and BLZ in the \twoMPZ catalog.
This identification will be useful to model the cross-correlation angular power spectrum of mAGN and BLZ, as described in Section \ref{sec:inter}.

Blazars are identified by cross-matching \twoMPZ with the \WIB catalog \cite{WIBRALS}. The latter is composed of radio-loud WISE sources detected in all four WISE filters, whose mid-infrared colors match typical colors of confirmed \g-ray emitting blazars.
We select mAGNs by cross-matching
%\footnote{The cross-match has been performed with CDS X-Match Service http://cdsxmatch.u-strasbg.fr/xmatch, searching for the closest galaxy within 5 arcsec} 
\twoMPZ with the AGN sample found in Ref.~\cite{Edelson2012}. The authors considered WISE and 2MASS data and defined a statistical discriminator by comparing the measured infrared colors, producing a complete sample of AGNs. This subset contains $\sim 10^4$ objects and we remove blazars obtained from the \WIB catalog.

\section{Measurements}
\label{sec:measur}

The cross-correlation APS is defined as:
\be 
C_\ell^{(ij)} = \frac{1}{2\ell+1} \sum_m a^{(i) \star}_{\ell m} a^{(j)}_{\ell m}
\ee
where: 
\be
a^{(i)}_{\ell m} = \int d\vec{n}\; \delta I^{(i)}(\vec{n}) \;Y_{\ell m} (\vec{n})
\ee
are the coefficients of the expansion of the fluctuations $\delta I^{i}(\vec{n})$ of the field $I^{i}(\vec{n})$ in terms of spherical harmonics $Y_{\ell m} (\vec{n})$. In our case $i$ and $j$ correspond to the \g-ray and galaxy map fields. We determine the APS with PolSpice\footnote{\href{http://www2.iap.fr/users/hivon/software/PolSpice/}{http://www2.iap.fr/users/hivon/software/PolSpice/}}, a public code that computes both the two-point angular cross-correlation function in real space and the APS. 
PolSpice is based on the fast spherical harmonic transforms allowed by isolatitude pixelisations and it corrects for the effects introduced by masking following the approach of Ref.~\cite{efstathiou04}. PolSpice also provides an estimate for the covariance matrix of the measurement, that will be used for the statistical analysis discussed in the following Sections.

Before computing the APS, we remove the monopole and dipole contributions from the input maps by applying the HEALPix routine \code{remove$\_$dipole}, in order to mitigate a possible leakage of these (large) terms to higher multipoles (an effect due to multipole-mixing introduced by the masks).

\begin{figure}[t]
  \centering
  \includegraphics[width=0.5\textwidth]{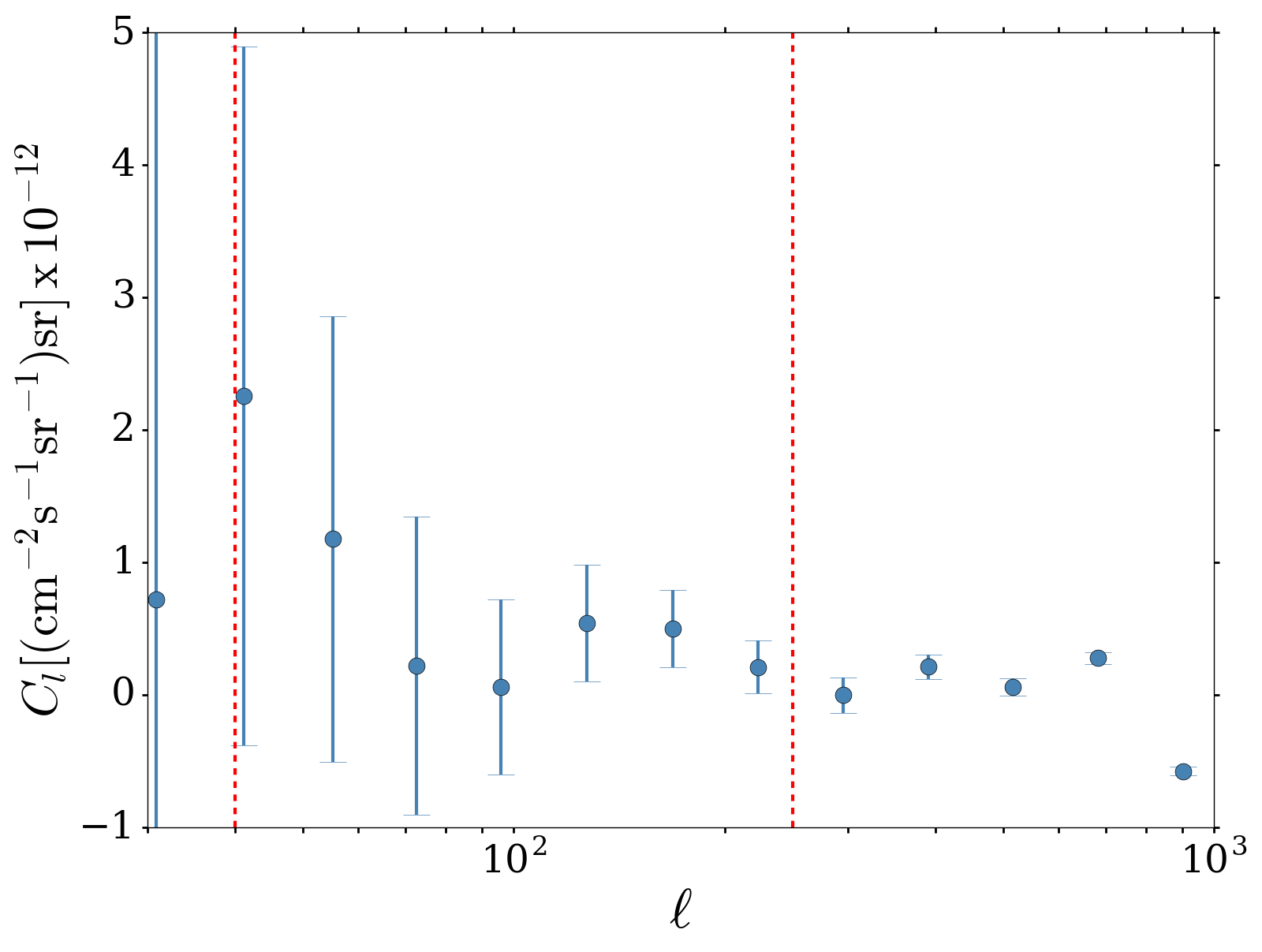}
  \caption{Measured APS between the \g-ray map in the $(1.2, 2.3)$ GeV energy bin and the complete 2MPZ map. The two dashed vertical lines bracket the multipole window over which the fits are performed. The lower limit is fixed to $\ell_{\rm} = 40$ while the upper bound $\ell_{\rm max}$ is determined from the beam window function and therefore depends depends on energy. All the measured APS can be retrieved at this link \cite{repository}.}
  \label{fig:cl_ex}
\end{figure}

The finite angular resolution of the {\it Fermi}-LAT instrument suppresses the angular power spectrum at high multipoles (the angular resolution of the galaxy surveys is significantly better than the {\it Fermi}-LAT one and the associated suppression would show up only at higher multipoles, in a range not considered in our analysis). In order to take this suppression into account, we correct the $C_\ell$ with the beam window function:
\be
W_\ell(E) = 2 \pi \int_{-1}^{1} d \cos\theta \, P_\ell(\cos\theta) \, {\rm PSF}(\theta, E)
\ee
where $P_\ell $ are the Legendre polynomials and ${\rm PSF}(\theta, E)$ denotes the {\it Fermi}-LAT\ point spread function for the specific  IRF and energy, as provided by the {\it Fermi} Science Tools. The energy-dependent  beam window function is averaged in each energy bin in accordance to the UGRB energy spectrum $E^{-\alpha}$, where the spectral index is taken at $\alpha = 2.3$ \cite{Ackermann:2014usa}:
\be
\langle W_\ell^k \rangle = \frac{\int_{E_{{\rm min}, k}}^{E_{{\rm max}, k}} W_\ell(E) E^{-\alpha} dE}{\int_{E_{{\rm min}, k}}^{E_{{\rm max}, k}} E^{-\alpha} dE}
\ee

The measured APS in the $k$-th energy bin is then defined as:
\be 
C^k_\ell = \frac{C^k_{\ell, {\rm raw}}}{\langle W^k_\ell \rangle {W_{\rm pix}}}\;,
\ee
where $C^k_{\ell, {\rm raw}}$ is the raw APS obtained from PolSpice in the $k$-th energy bin and
${W_{\rm pix}}$ is the pixel window function associated to the HEALPix pixeling.

For the analyses discussed in the next Section, we re-bin the measured APS in 15 evenly-spaced logarithmic multipole bins from 10 to 1000. Since the low multipoles $C^k_\ell$ can be affected by large-scale effects due to an imperfect Galactic foreground removal and at large multipoles $C^k_\ell$ by an imperfect PSF correction, especially when the beam window function starts deviating significantly from 1, we must identify a suitable multipole range over which we perform our analyses: the lower limit is conservatively set to $\ell_{\rm min} = 40 $; the upper limit $\ell_{max}$ is defined from the condition that the beam window function does not drop below a threshold corresponding approximately to the 68\% containment of the PSF in the specific $k$-th energy bin:
\be
\langle W_{\ell_{\rm max}}^k \rangle = 0.61,
\ee
or $l_{\rm max} =1000$, whichever is smaller. This condition makes $\ell_{\rm max}$ dependent on energy. The lower and upper bound of the multipole bins for each energy bin are shown in Table \ref{tab:enlist}.

\begin{table}
\centering
\begin{tabular}{|c|c|c|c|c|}
 \hline
Bin & $E_{\rm min}$ [GeV] & $E_{\rm max}$ [GeV]& $\ell_{\rm min}$ & $\ell_{\rm max}$ \\
 \hline
1 & 0.631 &1.202 &40 &220 \\
2 & 1.202 &2.290 &40 &250 \\
3 & 2.290 &4.786 &40 &307 \\
4 & 4.786 &9.120 &40 &487 \\
5 & 9.120 &17.38 &40 &695 \\
6 & 17.38 &36.31 &40 &907 \\
7 & 36.31 &69.18 &40 &1000 \\
8 & 69.18 &131.8 &40 &1000 \\
9 & 131.8 &275.4 &40 &1000 \\
10 & 275.4 &524.8 &40 &1000 \\
11 & 524.8 &1000.0 &40 &1000 \\
 \hline
\end{tabular}
\caption{Energy bins and their corresponding multipole ranges (identified with the procedure discussed in the text) over which our analysis is performed.}
\label{tab:enlist}
\end{table}

Fig.~\ref{fig:cl_ex} shows an example of the measured APS, in the $(1.2, 2.3)$ GeV energy bin, for the cross-correlation with the whole 2MPZ catalogue.  The plot also shows the multipole range $(l_{\rm min}, l_{\rm max})$ for this energy bin. Error bars are large at low multipoles because of cosmic variance, mask deconvolution and noise from Galactic foreground. They start becoming large also at multipoles above a few hundreds because of the size of the \Fermi-LAT PSF (and finite statistics).  

All the measured APS can be retrieved at this link \cite{repository}.

\begin{figure*}[t]
  \centering
  \includegraphics[width=0.95\textwidth]{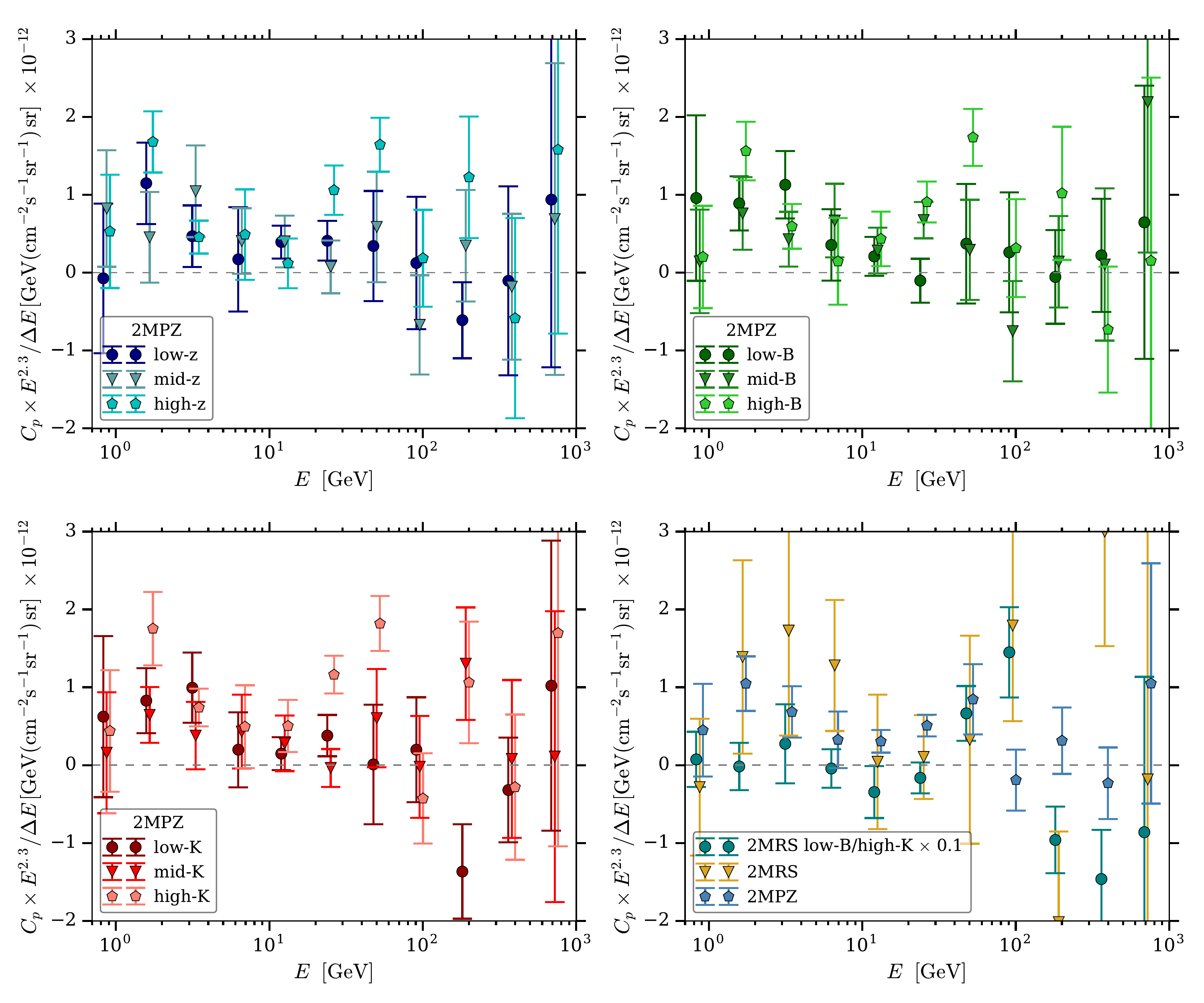}
  \caption{Multipole-independent APS $C_p^k$ as a function of the energy, for the different galaxy subsamples considered in this work.}
  \label{fig:cp}
\end{figure*}

\subsection{Amplitude and significance of the correlation}
\label{sec:ampli}

In order to provide a model-independent estimate of the amplitude and significance of the measured cross-correlations, we fit the APS in each energy bin with a term which is multipole-independent (i.e., a constant) that we call $C_p^k$.
This can be considered as the simplest model (i.e., a Poisson noise term) and provides an estimate of the amplitude which is similar to performing the average of the APS over the multipole range of interest.
A more refined treatment, which involves modeling of the \g-ray components in the unresolved sky, is presented in the next section.

Fig.~\ref{fig:cp} shows $C_p^k$ as a function of the energy bin for the cross-correlation of the \g-ray flux maps with each of our galaxy subsamples. The plot indicates the presence of a correlation signal between the galaxy distribution and \g-rays for all subsamples. In fact, the $C_p^k$ are systematically positive (i.e., they do not fluctuate around zero) and deviate from a null signal. To assess the significance of the measurements, we compare the $\chi^2$ of a null signal with the $\chi^2$ obtained from the $C_p^k$ fit. 

We adopt a $\chi^2$ estimator defined as:
\begin{eqnarray}
 &&\chi^2 = \label{eq:chi2}  
 \sum_{k=1}^{11}\sum_{\Delta\ell, \Delta\ell'} 
\\ &&(C_{\Delta\ell}^{k, {\rm mod}} - C_{\Delta\ell}^{k, {\rm exp}})\Gamma^{-1}_{\Delta\ell,\Delta\ell',k}(C_{\Delta\ell'}^{k, {\rm mod}}-C_{\Delta\ell'}^{k, {\rm exp}})\;,\nonumber
\end{eqnarray}
where $C_{\Delta\ell}^{k, {\rm exp}}$ is the measured APS in the energy bin $k$ and multipole bin $\Delta\ell$, $C_{\Delta\ell}^{k, {\rm mod}}$ is the APS model and $\Gamma_{\Delta\ell,\Delta\ell',k}$ is the covariance matrix, obtained from the PolSpice covariance through multipole re-binning. We neglect the covariance between different energy bins since the main source of error comes from the Poisson noise of the \g-ray maps, something which exhibits no correlation among different energy bins. Eq.~\ref{eq:chi2} will be adopted throughout the paper for model comparison, including the analysis in terms of \g-ray modeling as discussed in the next sections.

We define a $\chi^2$ difference $\Delta \chi^2 = \chi^2_{\rm null} - \chi^2_{Cp}$, where $\chi^2_{\rm null}$ is the null signal obtained from Eq.~\ref{eq:chi2} by using $C_{\Delta\ell, {\rm mod}}^k = 0$, and $\chi^2_{C_p}$ is obtained using $C_{\Delta\ell,{\rm mod}}^k = C_p^k$. Table \ref{tab:chi2poiss} shows the results for the different subsamples: for each case, $\Delta \chi^2 >0$ with values ranging from 3 to 29. We postpone comments about the variation of the significance across different galaxy samples, since the simple constant APS model adopted for this part of the analysis may be more suited for some subsamples than for others. More physically motivated models and their significances will be discussed in the next section. We simply note here that Table~\ref{tab:chi2poiss} shows a general significant deviation from the null hypothesis. 

Since in our correlation measurement we employ maps of the integrated \g-ray flux, we expect the energy spectrum to follow the integrated energy spectrum of the UGRB, namely $I_{\rm UGRB}=\int_{\Delta E} dE\, dI_{\rm UGRB}/dE$. By multiplying the vertical axis in Fig.~\ref{fig:cp} by $E^2/\Delta E$, we show (approximately) the differential energy spectrum of the \g-ray emission responsible for the correlation signal, rescaled by $E^{-2}$. The statistical significance is not enough to derive firm conclusions on the energy dependence, but different subsets seem to indicate a \g-ray population with a spectral index close to $-2$ (so with a flat spectrum in Fig.~\ref{fig:cp}) at high energy, while a source with a softer spectrum at low energy.
Among the astrophysical \g-ray emitters, blazars typically show a hard spectrum (with index about $-2$), while other types of AGN (i.e., misaligned) and star forming galaxies have softer emission.

\begin{table}[t]
\centering
\begin{tabular}
{|c|c|c|c|}
 \hline
~~~~Subset~~~~ & $~\chi^2_{\rm null}$~ & ~$\chi^2_{C_p}~$ & ~$\Delta \chi^2$~ \\
 \hline
2MPZ (full) &90.6 &76.0 &14.6\\ 
2MRS (full) &53.7 &49.6 &4.1\\ 
Lowz &67.8 &64.7 &3.1\\ 
Mid-$z$ &81.0 &74.3 &6.7\\ 
High -$z$ &92.9 &72.3 &20.5\\ 
Low-K &72.5 &65.9 &6.5\\ 
Mid-K &77.1 &72.3 &4.7\\ 
High-K &107.1 &78.5 &28.6\\ 
Low-B &70.8 &65.1 &5.7\\ 
Mid-B &80.4 &71.2 &9.3\\ 
High-B &104.0 &81.1 &23.0\\ 
 \hline
\end{tabular}
\caption{Comparison of the best-fit $\chi^2$ results for the no-signal case and a multipole-independent $C_p^k$. The total number of data-points considered for each sample (including energy and multipole bins) is 114.}
\label{tab:chi2poiss}
\end{table}

\section{Results and Interpretation}
\label{sec:inter}
\subsection{Models}
We model the clustering of galaxies in the samples presented in Sec.~\ref{subsec:galsub} and of \g-ray emitters mentioned in the introduction by making use of the halo model approach. Galaxies are assumed to follow the matter power spectrum with matter distributed in halos, and with the number of galaxies per halo defined by the so-called halo occupation distribution (HOD). The latter has been derived by fitting the auto-correlation APS of galaxies, as described in the Appendix.

\begin{figure}[t]
\vspace{-4.cm}
\centering
  \includegraphics[width=0.55\textwidth]{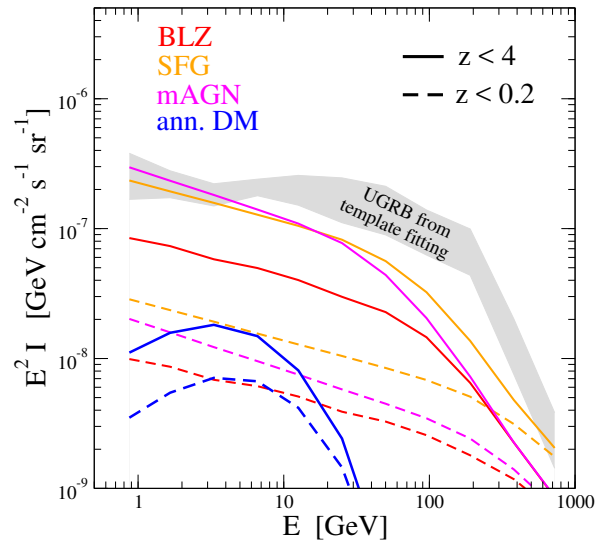}
\vspace{-0.5cm}
  \caption{Energy spectrum of the UGRB as determined from the method described in Section~\ref{sec:fermidata} (gray band) together with the predicted contribution for the reference model of the different \g-ray emitters considered in this work. With dashed lines, we show also their contribution to the UGRB from $z<0.2$ (approximately the range of redshift considered in this work).}
  \label{fig:ugrb}
\end{figure}

The cross-correlation APS with \g-ray sources is computed as in Ref.~\cite{2015ApJS..221...29C} with two differences.
The first involves the combination of flat spectrum radio quasars and BL Lacs into a single (effective) blazar class, as done in Ref.~\cite{ajello2015IGRB}. The contribution to the UGRB of the \g-ray emitters considered in this work is shown in Fig.~\ref{fig:ugrb}. Note from the figure that we expect the \g-ray emission we analyze in this paper (which is produced at $z<0.2$) to amount to a small fraction of the total UGRB, roughly around 10\% at low energy. The three classes of emitters provide comparable contributions, within a factor $\mathcal{O}(1)$.

The second difference is related to the modeling of the shot-noise term.
This contribution to the APS is the part at zero angular separation (i.e., $\ell$-independent) of the 1-halo term. As recognized in Ref.~\cite{2015ApJS..221...29C}, its modeling can be very delicate. Here, we do not attempt to include it in our halo-modeling, whilst we follow two data-driven approaches.
In the simplest approach, we just fit the shot-noise contribution (which is a constant term for each APS) by assuming a power-law energy dependence: 
\be
\frac{d C_p^{(j)}}{d E} = N_j E^{-\alpha_j}\;,
\label{eq:cp}
\ee
where the index $j$ labels the galaxy sample.
In this way, we introduce 2 parameters (normalization and power-law index) for each galaxy sample. This approach will be called the ``\cpfree'' fit.

The second approach, which is our reference one (called \std), uses the fact that the shot-noise is given by the average of the \g-ray flux of all the galaxies $N_{\rm gal}$ of a given sample $j$ (let us remember that index $k$ denotes the energy bin):
\be
C_{p}^{(jk)}=\frac{1}{N_{\rm gal}^j}\sum_{i=0}^{N_{\rm gal}^j}\,F^\gamma_{\Delta E_k,j}\;.
\ee
We adopt some empirical relations to predict $F^\gamma_{\Delta E_k,j}$ from the optical/infrared magnitude of the each galaxy in the catalog, as described in the Appendix. Furthermore, we need to identify blazars and misaligned AGNs in the \twoMPZ catalog, for which we use the procedure outlined in Sec.~\ref{subsec:galsub}. With these two ingredients, we are able to estimate the shot-noise contribution.

The signal associated to annihilating DM is computed, again following the halo model approach, as described in Ref.~\cite{Regis:2015zka}, with the ``boost-factor'' taken from Ref.~\cite{Moline:2016pbm}. The contribution depends on two parameters, the particle DM mass $M_\chi$ and the annihilation cross section $\sigma v$. We will consider four different DM models, referring to four specific annihilation final states endowed with different spectra and representative for a typical WIMP DM: $b\bar b$, $\tau^+\tau^-$, $W^+W^-$, and $\mu^+\mu^-$.

Summarizing, we fit the cross-correlation APS of the galaxy samples presented in Sec.~\ref{subsec:galsub} and the \Fermi-LAT \g-rays intensity maps with two approaches:
\begin{itemize}
\item {\sl Reference} model:
\begin{align*}
& C_{\ell, {\rm mod}}^{(jk)} = C_{\ell, {\rm DM}}^{(jk)}(M_\chi, \sigma v) \\
& + N_{\rm SFG} \times (C_{\ell, {\rm SFG}}^{(jk)} + C_{p, {\rm SFG}}^{(jk)}) \\ 
& + N_{\rm BLZ} \times (C_{\ell, {\rm BLZ}}^{(jk)} + C_{p, {\rm BLZ}}^{(jk)}) \\
& + N_{\rm mAGN} \times (C_{\ell, {\rm mAGN}}^{(jk)} + C_{p, {\rm mAGN}}^{(jk)}). 
\end{align*}
In this approach, the total number of free parameters is 5, i.e., 3 normalizations ($N_{\rm SFG}$, $N_{\rm BLZ}$ and $N_{\rm mAGN}$) for the astrophysical contributions and 2 terms for the annihilating DM contribution ($M_\chi$ and $\sigma v$). The annihilation rate will be expressed in terms of the ``thermal'' (or ``natural scale'') value $\langle\sigma v\rangle_{\rm th} = 3\times 10^{-26} {\rm cm^3/s}$ by trading it for a dimensionless parameter $N_{\rm DM} = \sigma v/ \langle\sigma_a v\rangle_{\rm th}$.
\item {\sl Free} $C_p$ model:
\begin{align*}
& C_{\ell,{\rm mod}}^{(jk)} =C_{\ell, {\rm DM}}^{(jk)}(M_\chi, \sigma v) + N_{\rm SFG} \times C_{\ell, {\rm SFG}}^{(jk)} \\ 
& + N_{\rm BLZ} \times C_{\ell, {\rm BLZ}}^{(jk)} + N_{\rm mAGN} \times C_{\ell, {\rm mAGN}}^{(jk)} + C^{(jk)}_p,
\end{align*}
where the last term $C^{(jk)}_p = \int_{E_{\rm min}^k}^{E_{\rm max}^k}dE\, dC_p^{(j)}/dE$, with ${E^k_{\rm min}}$ and ${E^k_{\rm max}}$ being the energy boundaries of the $k$-th energy bin.
With respect to the previous case, this model adds 2 parameters for each sample $j$, associated with the \cp\ term (see Eq.~\ref{eq:cp}). 
\end{itemize}

\subsection{Statistical analysis}

Our fit is performed with the Monte Carlo parameter estimation code CosmoSIS \cite{Zuntz:2014csq}. Since the order of magnitude of each parameter is unknown, we use a Metropolis-Hastings sampler with a flat prior in log-scale for each parameter.

The galaxy subsamples listed in Sec.~\ref{subsec:galsub} are analyzed separately. For the cases involving three bins (redshift, B-luminosity and K-luminosity), we fit simultaneously the  APS of the different bins, which are independent from each other (since the galaxy subsamples are not overlapping). For these samples, the number of parameters in the fit is 5 (11) in the \std\ (\cpfree) model. For all the other samples the number of parameter is 5 (7) in the \std\ (\cpfree) model. 

\begin{figure*}[t]
  \centering
  \includegraphics[width=0.9\textwidth]{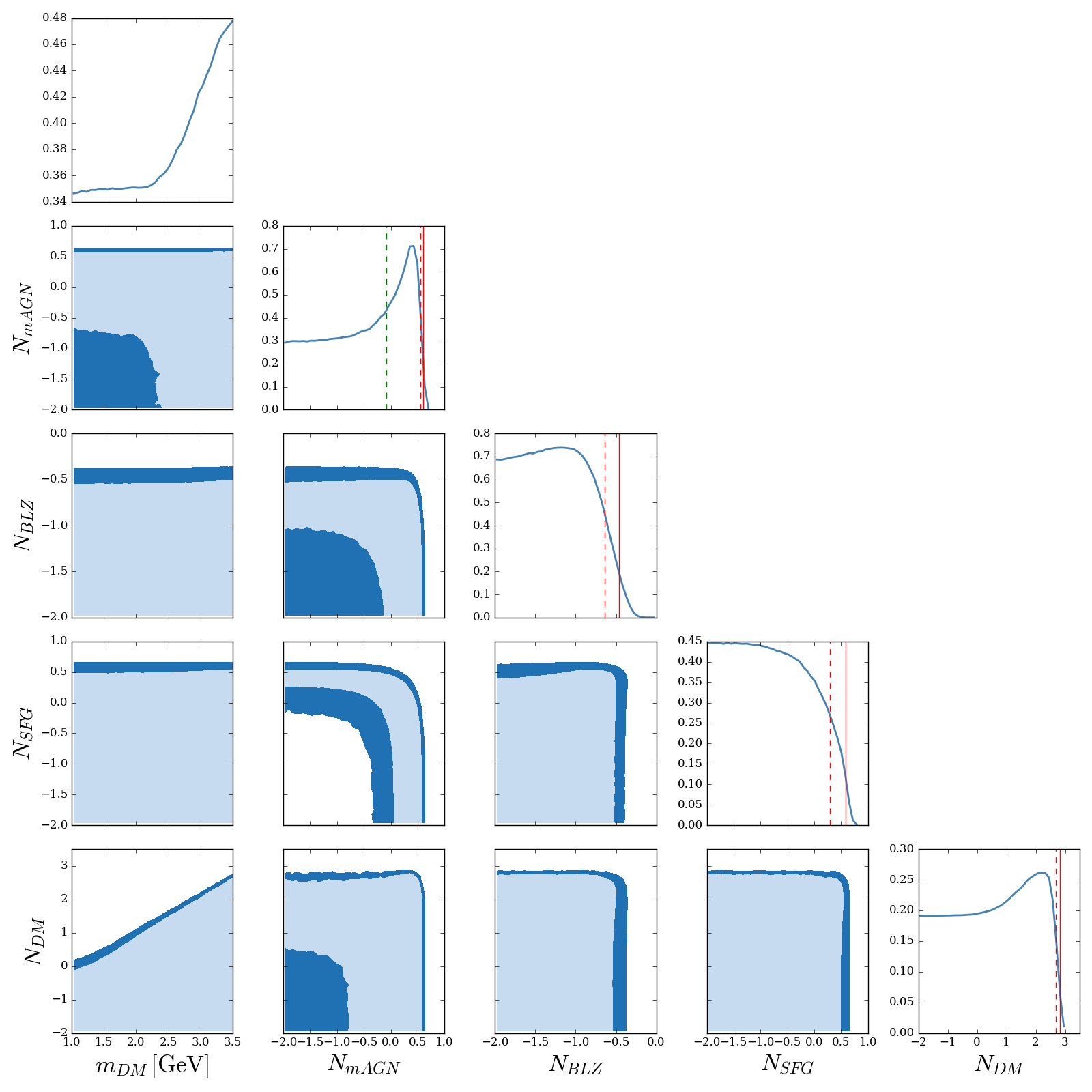}
  \caption{Fit results for the 2MPZ redshift slicing subset for the {\sl reference} analysis. All parameters are shown in log-scale. The vertical dashed and solid red (green) lines denote the 68\% and the 95\% CL upper (lower) limits obtained from the profile likelihood, respectively. In the 2D plots, the 68\% regions are identified in cyan while the 95\% regions are in dark blue. The 1D profile likelihood distributions on the diagonal are individually normalized to unity.}
  \label{fig:triangleplot}
\end{figure*}

As an example of the outcome, in Fig.~\ref{fig:triangleplot}, we show the triangular plot obtained by fitting the 2MPZ split into redshift bins. The vertical dashed and solid red (green) lines denote the 68\% and the 95\% CL upper (lower) limits found with the profile likelihood, respectively. In the 2D panels, the 68\% regions are identified in cyan while the 95\% regions are in dark blue. 
In this example, the only normalization which is not compatible with zero (at $1\sigma$ level) is for the mAGN population, see last panel of second row.

All the triangular plots for the various cases are available at this link \cite{repository}. In the following, for the sake of brevity, we will focus our discussion on the 1D profile likelihood distributions, except in the case of the DM parameters, for which we will discuss also the 2D plane showing the bounds on the particle DM parameters in the canonical annihilation rate vs DM mass space. 

\begin{figure*}[t]
  \centering
  \includegraphics[width=0.95\textwidth]{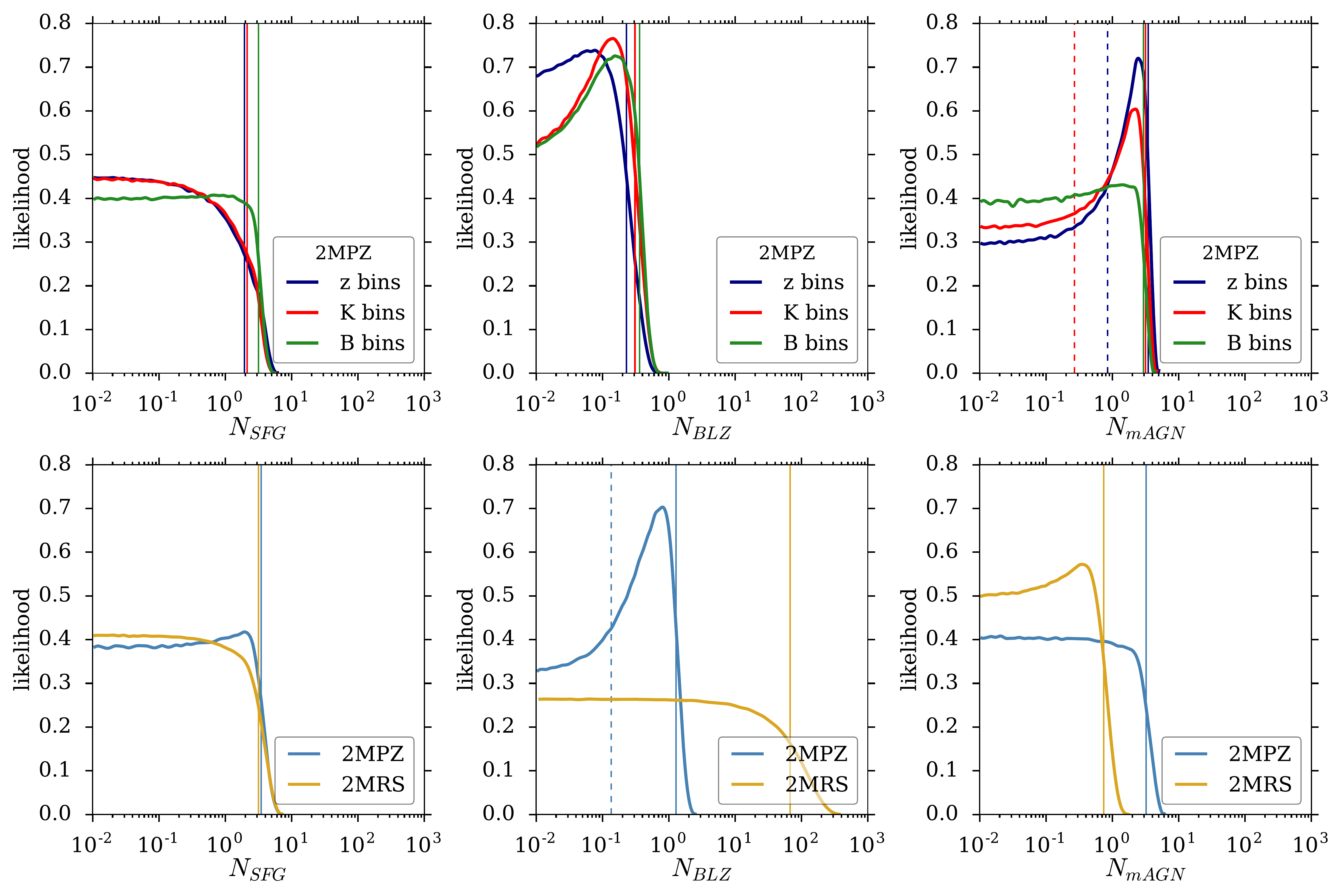}
  \caption{Profile likelihood distributions for the normalization parameters of the astrophysical \g-rays components, for the \std\ analysis.
The upper panels show the results obtained for the different subsamples of the 2MPZ catalog. The lower panels show the results for the full 2MPZ and for the low-redshift 2MRS catalogs. The vertical solid (dashed) lines indicate the 68\% upper (lower) limits (whenever present in the plots).}
  \label{fig:plot1D}
\end{figure*}

Fig.~\ref{fig:plot1D} summarizes the results on the normalization parameters of the astrophysical \g-ray sources for the \std\ analysis. The upper three panels shows the 1D likelihood distributions for SFG, BLZ and mAGN obtained by organizing the galaxy data into the three different subsamples that differentiate the galaxies in terms of redshift, K luminosity and B luminosity. The lower three panels show the results for SFG, BLZ and mAGN when the full 2MPZ catalog (blue) or the low-redshift 2MRS catalog (yellow) are used. 

\begin{figure*}[t]
  \centering
  \includegraphics[width=0.95\textwidth]{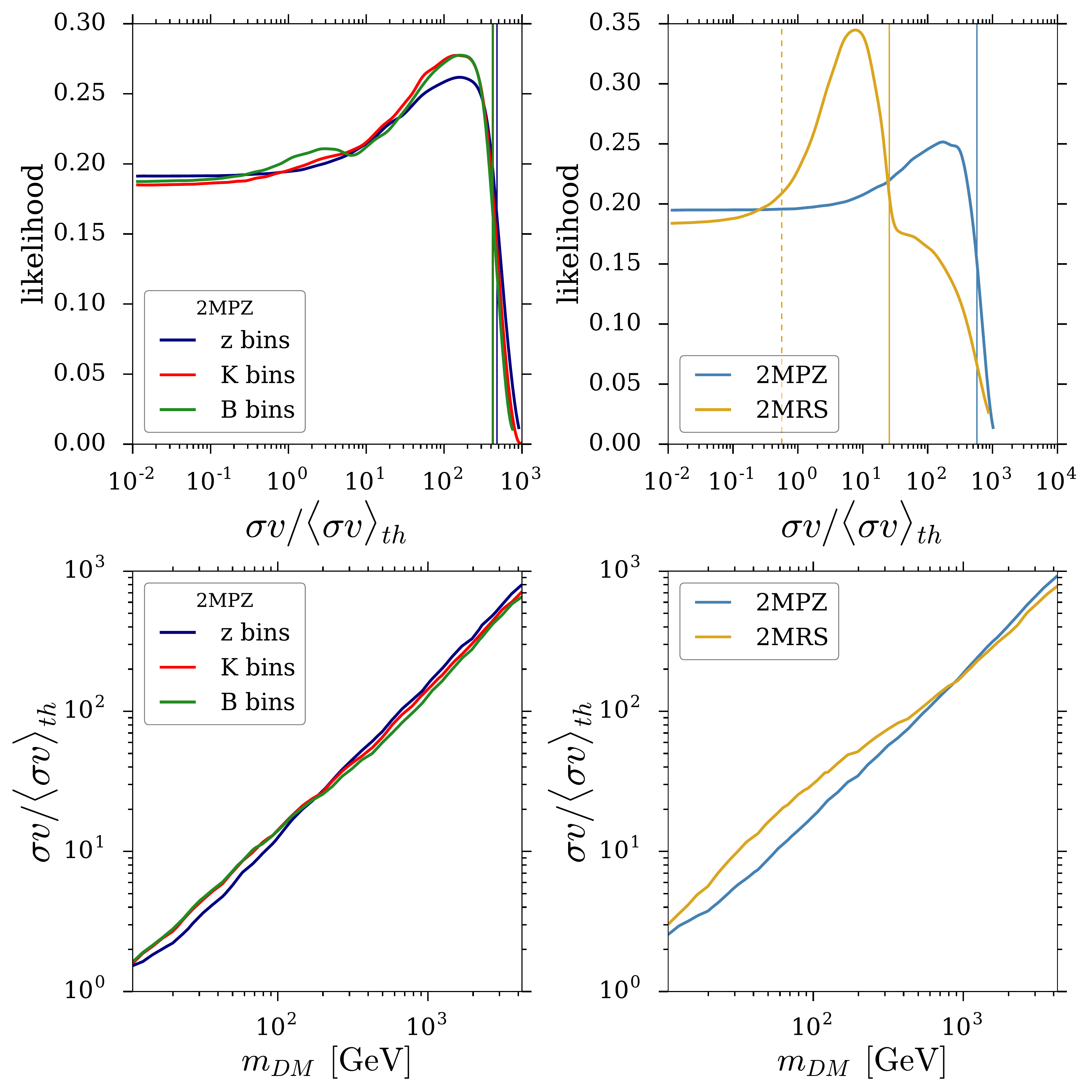}
  \caption{Results for the DM case obtained with the the \std\ analysis and $b\bar b$ annihilation channel (with ``boost-factor'' from Ref.~\cite{Moline:2016pbm}). The upper panels show the profile likelihood distribution for the annihilation rate. The lower panels show the 95\% CL upper bounds for the annihilation rate vs the DM mass. The two panels in the first column refer to the analyses performed on different organization of the galaxy samples (redshift, K luminosity and B luminosity). The two panels in the second column refer to the analyses on the full 2MPZ catalog and on the low-redshift 2MRS catalog.}
  \label{fig:plot2D}
\end{figure*}

\begin{figure*}[t]
  \centering
  \includegraphics[width=0.95\textwidth]{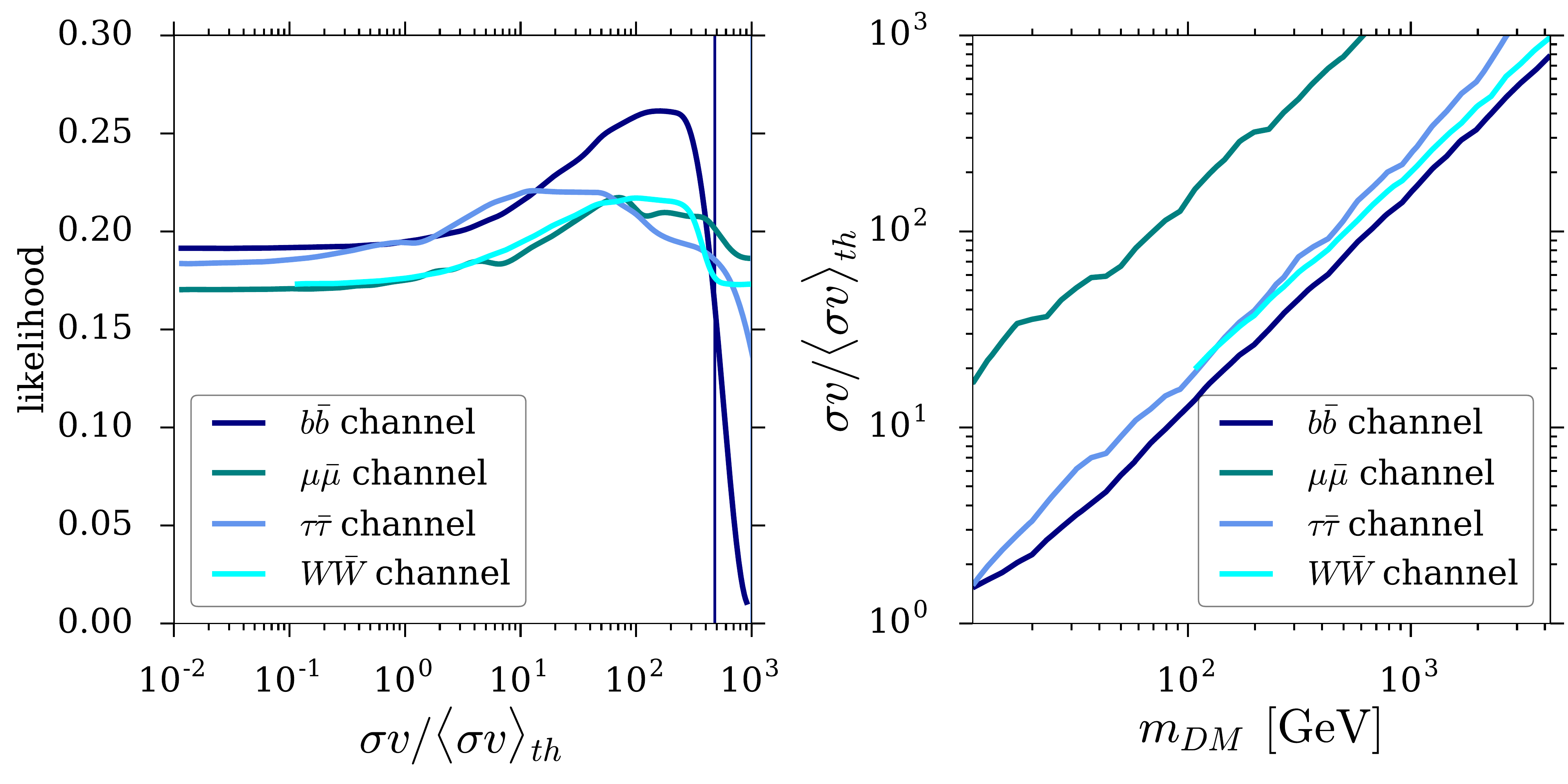}
  \caption{Left panel: profile likelihood for the DM annihilation rate for the four annihilation channels considered in this analysis. Right panel: 95\% CL upper bounds on the DM annihilation cross-section as a function of the DM mass for the same annihilation channels. The plot refers to the \std\ analysis performed combining the three $z$-bins of the \twompz\ catalog.}    
  \label{fig:plot2Dch}
\end{figure*}

The corresponding DM results for the \std\ analysis are shown in Fig.~\ref{fig:plot2D}, for DM annihilating in the $b\bar b$ channel. The upper panels show the likelihood distributions for the annihilation rate for the different galaxy subsamples, while the lower panels show the corresponding 95\% CL bounds on the annihilation rate as a function of the DM mass for the same annihilation channel. The bounds for all the four annihilation channels considered in this work ($b\bar b$, $\tau^+\tau^-$, $W^+W^-$, and $\mu^+\mu^-$) and for the analysis performed combining the three $z$-bins of the \twompz\ catalog are shown in Fig.~\ref{fig:plot2Dch}. 
This figure can be considered as the summary plot for what concerns the bounds on WIMP DM derived in this work.

\begin{figure*}[t]
  \centering
  \includegraphics[width=0.95\textwidth]{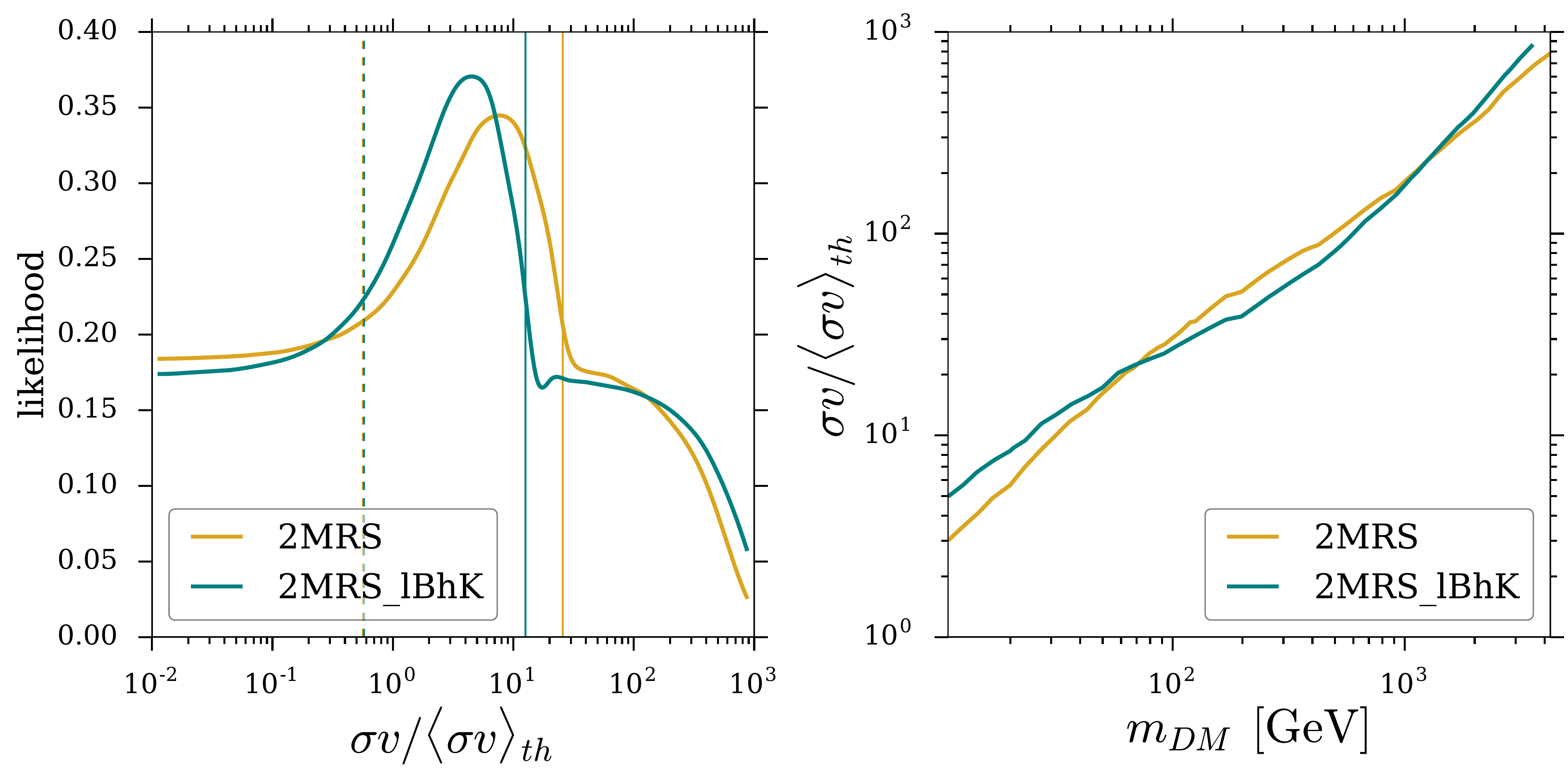}
  \caption{Results for the DM case using galaxy catalog samples expected to be more sensitive to the DM \g-ray signal, i.e., the low redshift catalog 2MRS (yellow line) and its combination with the High K--Low B subsample (blue line). The results refer to the \std\ analysis. Left: profile likelihood distributions for the DM annihilation rate; the vertical solid (dashed) lines indicate the 68\% upper (lower) limits (whenever present in the plots). Right: 95\% upper bound on the annihilation rate vs the DM mass.
} 
  \label{fig:plot1D_lBhK}
\end{figure*}

As a further investigation of the DM case, Fig.~\ref{fig:plot1D_lBhK} considers galaxy samples for which the cross-correlation with \g-rays is expected to be enhanced, i.e., the low-redshift 2MRS sample and its combination with the High K--Low B subsample of the 2MPZ catalog. Again, the left and right panels show the likelihood distribution for the annihilation rate and the 95\% CL bounds in the annihilation rate vs mass plane. 

Table~\ref{tab:bf} reports the best-fit values and the 68\% upper and lower bounds (whenever present) for the astrophysical and DM parameters, for the different galaxy samples. Discussion and interpretation of the results are presented in the next section.

\begin{figure*}[t]
  \centering
  \includegraphics[width=0.95\textwidth]{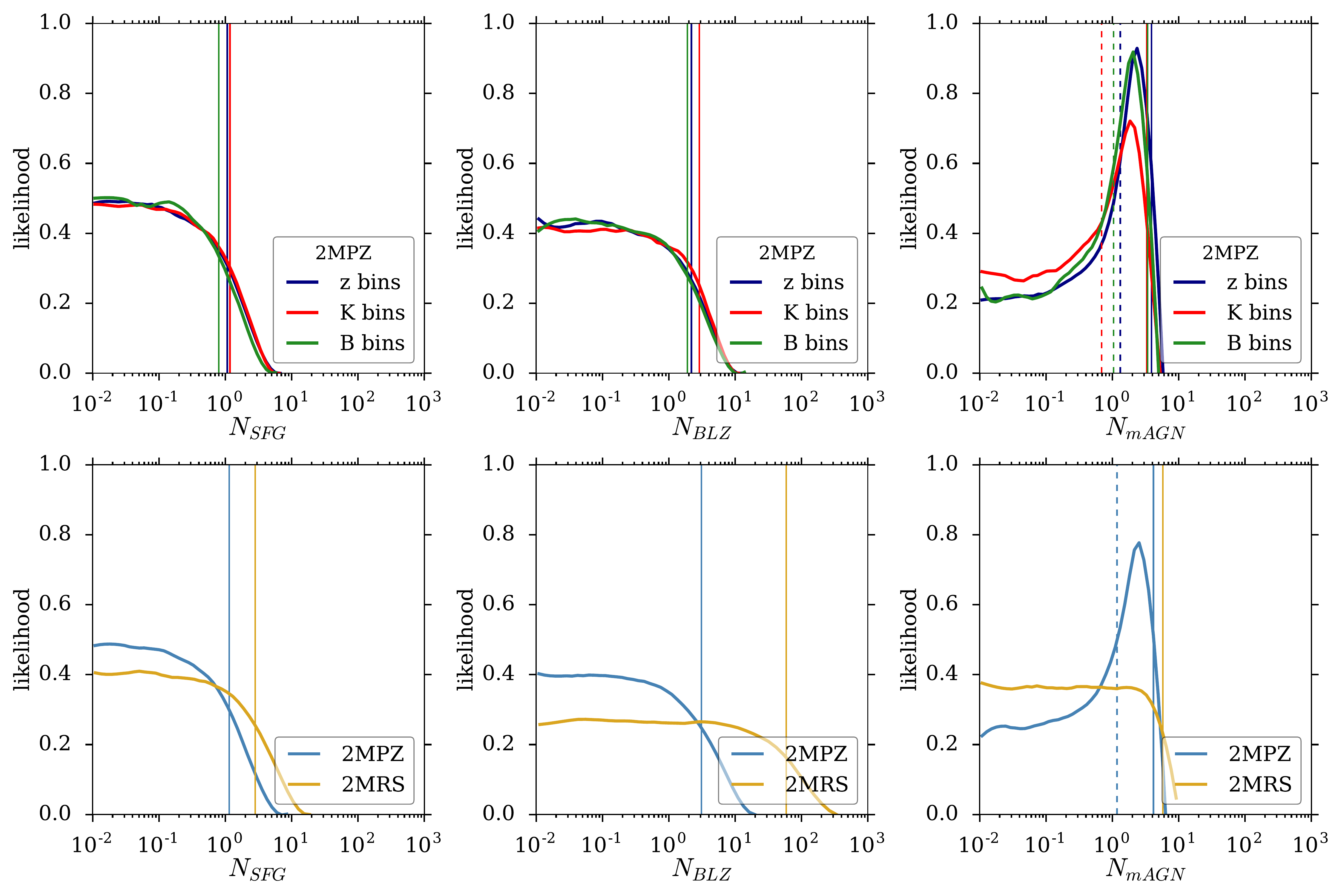}
  \caption{The same as Fig.~\ref{fig:plot1D}, but for the \cpfree\ analysis. 
}
  \label{fig:plot1Dcpfree}
\end{figure*}

\begin{figure*}[t]
  \centering
  \includegraphics[width=0.95\textwidth]{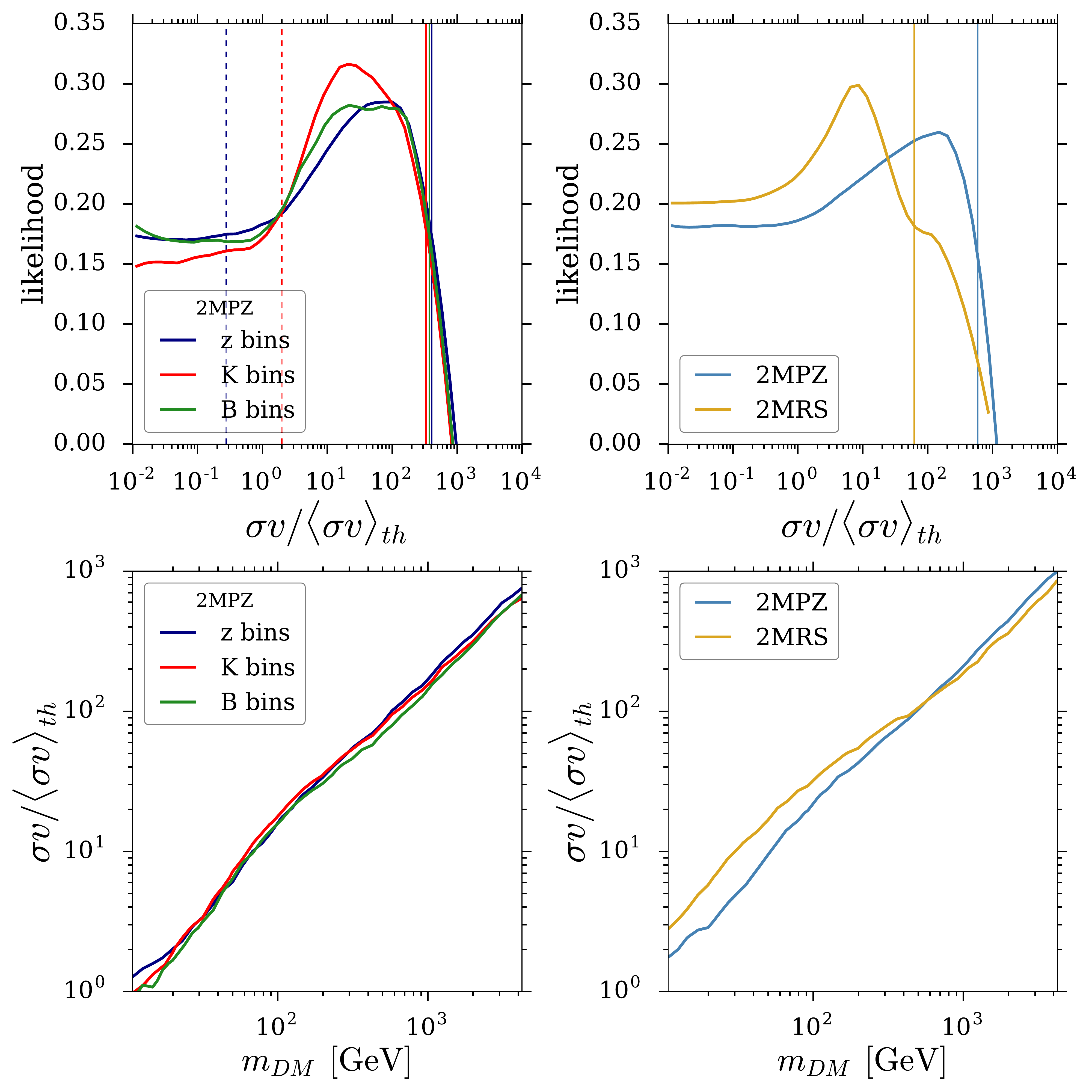}
  \caption{Same as Fig.~\ref{fig:plot2D} but for the \cpfree\ analysis. 
}
  \label{fig:plot2Dcpfree}
\end{figure*}

For the \cpfree\ analysis, the results are shown in Figs.~\ref{fig:plot1Dcpfree} and \ref{fig:plot2Dcpfree}, that mirror the  information in Figs.~\ref{fig:plot1D} and \ref{fig:plot2D}, respectively. Table~\ref{tab:bfcpfree} lists the best-fit values and the 68\% upper and lower bounds (whenever present) for the astrophysical and DM parameters, for the different galaxy samples. In Table \ref{tab:bfcppar} we show the best fit results for the $C_p$ normalizations and power-law indexes.

Finally, the statistical significance of the {\sl reference} and \cpfree\ models as compared to the null hypothesis of absence of signal are shown in Table~\ref{tab:chi2} in terms of the $\chi^2$ differences.

\begin{table*}[t]
\centering
\begin{tabular}{|c|ccc|ccc|ccc|ccc|}
 \hline
Sample  & \multicolumn{3}{|c} {$N_{\rm mAGN}$} & \multicolumn{3}{|c} {$N_{\rm SFG}$} & \multicolumn{3}{|c} {$N_{\rm BLZ}$} & \multicolumn{3}{|c|} {$N_{\rm DM}$} \\
~& BF &low&up& BF &low&up& BF &low&up& BF &low&up\\
\hline
2MPZ (full) & 0.02 & - & 3.24 & 0.76 & 0.13 & 1.29 & 1.95 & - & 3.47 & 190.55 & - & 575.44 \\ 
2MRS (full) & 0.35 & - & 0.74 & 0.06 & - & 67.61 & 0.02 & - & 3.16 & 7.59 & 0.56 & 25.70 \\ 
$z$ bins & 2.45 & 0.85 & 3.47 & 0.07 & - & 0.23 & 0.02 & - & 1.95 & 181.97 & - & 478.63 \\ 
B bins & 1.45 & - & 2.95 & 0.15 & - & 0.36 & 0.66 & - & 3.16 & 165.96 & - & 416.87 \\ 
K bins & 2.09 & 0.27 & 3.16 & 0.14 & - & 0.31 & 0.03 & - & 2.14 & 165.96 & - & 426.58 \\ 
 \hline
\end{tabular}
\caption{Best fit and 68\% C.L. interval of the various parameters in the fit for the {\sl reference} case. When the lower bound is not reported, it means that it is compatible with zero at the quoted CL.}
\label{tab:bf}
\end{table*}

\begin{table*}[t]
\centering
\begin{tabular}{|c|ccc|ccc|ccc|ccc|}
 \hline
Sample  & \multicolumn{3}{|c} {$N_{\rm mAGN}$} & \multicolumn{3}{|c} {$N_{\rm SFG}$} & \multicolumn{3}{|c} {$N_{\rm BLZ}$} & \multicolumn{3}{|c|} {$N_{\rm DM}$} \\
~& BF &low&up& BF &low&up& BF &low&up& BF &low&up\\
\hline
2MPZ (full) & 3.02 & 1.17 & 4.17 & 0.01 & - & 3.09 & 0.02 & - & 1.15 & 120.23 & - & 588.84 \\ 
2MRS (full) & 0.49 & - & 5.75 & 0.09 & - & 58.88 & 0.07 & - & 2.82 & 11.22 & - & 61.66 \\ 
$z$ bins & 2.63 & 1.32 & 3.89 & 0.01 & - & 2.19 & 0.04 & - & 1.07 & 77.62 & 0.28 & 407.386\\ 
B bins & 2.45 & 1.05 & 3.39 & 0.05 & - & 1.91 & 0.02 & - & 0.79 & 25.70 & - & 371.54\\ 
K bins & 2.24 & 0.69 & 3.31 & 0.07 & - & 2.88 & 0.02 & - & 1.17 & 19.05 & 2.00 & 331.13 \\ 
 \hline
\end{tabular}
\caption{Best fit and 68\% C.L. interval of the various parameters in the fit for the \cpfree case. When the lower bound is not reported, it means that it is compatible with zero at the quoted CL.}
\label{tab:bfcpfree}
\end{table*}

\begin{table*}[t]
\centering
\begin{tabular}{|c|c|c|c|c|c|c|}
 \hline
~~~~Sample~~~~ & $N_0$ & $\alpha_0$ & $N_1$ & $\alpha_1$ & $N_2$ & $\alpha_2$ \\
\hline
2MPZ (full) &$9.77 \times 10^{-14}$ &~0.48 & & & &\\ 
2MRS (full) &$5.37 \times 10^{-14}$ &~0.61 & & & &\\ 
$z$ bins &$3.98 \times 10^{-16}$ &$-0.96$ &$2.51 \,\, 10^{-17}$ &$-0.29$ &$9.55 \times 10^{-14}$ &~0.58\\ 
B bins &$6.92 \times 10^{-19}$ &~0.28 &$2.57 \,\, 10^{-15}$ &$-0.04$ &$1.00 \times 10^{-13}$ &~0.54\\ 
K bins &$1.95 \times 10^{-15}$ &~0.49 &$2.24 \times 10^{-14}$ &~0.61 &$8.91 \times 10^{-14}$ &~0.61\\ 
 \hline
\end{tabular}
\caption{Best fit of the shot-noise parameters of Eq.(\ref{eq:cp}) for the \cpfree\  case.}
\label{tab:bfcppar}
\end{table*}

\begin{table*}[t]
\centering
\begin{tabular}{ |c|c|c|c|c|c|  }
 \hline
Subset & $\chi^2_{\rm null}$ & $\chi^2_{\rm ref}$ & $\chi^2_{{\rm free}\; C_p}$ & $\Delta  \chi^2_{\rm null - ref}$ & $\Delta \chi^2_{\rm null - free \; C_p}$ \\
 \hline
2MPZ (full) &95.63 &76.62 &78.58 &19.01 &17.05\\ 
2MRS (full) &58.75 &55.3 &55.98 &3.45 &2.77\\ 
$z$ bins &253.07 &229.24 &230.36 &23.83 &22.71\\ 
B bins &263.21 &233.82 &236.18 &29.39 &27.03\\ 
K bins &270.44 &241.32 &242.98 &29.12 &27.46\\ 
 \hline
\end{tabular}
\caption{Comparison of the best-fit $\chi^2$ results for the absence of signal ($\chi^2_{\rm null}$), the {\sl reference} analysis ($\chi^2_{\rm ref}$), the \cpfree\ analysis ($\chi^2_{{\rm free}\; C_p}$) and the relative differences of the two latter with the no-signal case.}
\label{tab:chi2}
\end{table*}

\subsection{Interpretation of the results}

\label{sec:disc}

In this Section we discuss the interpretations of the results presented in the previous section and conclude the consequences for the extragalactic \g-ray populations considered in our modeling.

\subsubsection{Star forming galaxies}
\label{sec:sfg}
Star-forming galaxies are poorly constrained by our analysis.
We find no relevant peak in the 1D likelihood distributions (i.e., in the left panels of Figs.~\ref{fig:plot1D} and \ref{fig:plot1Dcpfree}) and upper bounds on its normalization are around $2$--$3$ times the reference model. This implies that in order to provide a significant contribution to the cross correlation APS measurements derived in this work, SFG would overshoot the total UGRB intensity level shown in Fig.~\ref{fig:ugrb}.
In other words, we found that SFG are a subdominant component of the UGRB at low redshift.

\subsubsection{Misaligned AGNs}
\label{sec:magn}
Misaligned AGNs appear to be the population which can explain the bulk of the measured signal.
They are the only population that is singled out with statistical confidence in several datasets.
It is interesting to note the power of the ``tomographic'' approach. As shown in the bottom right panel of Fig.~\ref{fig:plot1D}, when considering the full \twoMPZ sample, no peak is present in the likelihood distribution. However, the evidence appears when considering the $z$ and $K$-luminosity bins (top right panel). 
A preference for mAGNs is also found in the \cpfree\ case, as shown in Fig.~\ref{fig:plot1Dcpfree}. Note that in this case, the mAGN normalization exhibits a lower limit already in the full \twoMPZ sample.

We see that the \twoMRS catalog seems to set an upper bound for the mAGN normalization that excludes the best-fits obtained with all the other samples (see bottom right panel of Fig.~\ref{fig:plot1D}). On the other hand, this does not happen in the \cpfree\ case (bottom right panel of Fig.~\ref{fig:plot1Dcpfree}), where instead the upper limit is consistent with the normalizations estimated from other galaxy samples. We remind that in the latter case, the $C_p$ are allowed to vary and are determined by the fit. These facts point toward a possible overestimate of the mAGN shot-noise at very low-$z$ (i.e., in the range covered by \twoMRS) in the \std\ model. In fact, in this case the shot noise has been derived from relations which show significant scatter (see Appendix): when applied to a very small volume like in the case of 2MRS, the shot-noise estimate might be not very accurate.
A dedicated analysis focusing on the low-redshift \twoMRS catalog will be the subject of future work.

\subsubsection{Blazars}
\label{sec:blz}
In the analysis of blazars, we can appreciate again how the tomographic approach tightens the bounds in Fig.~\ref{fig:plot1D}, pushing the normalization to lower values when going from the lower to the upper panel. Taken at face value, the results of the \std\ case (reported also in Table~\ref{tab:bf}) would indicate that BLZ are constrained to be a subdominant component of the total UGRB (see Fig.~\ref{fig:ugrb} where the BLZ component should be rescaled by a factor $N_{BLZ}$). Blazars are so constrained essentially because of their large shot-noise term that contributes in a non-negligible way to the cross APS signal we measure.

On the other hand, in the \cpfree\ model, the bounds become weaker, actually suggesting the opposite picture, namely that BLZ are a subdominant component of the cross-correlation we measure and of the UGRB at low redshift. Indeed, they need $N_{BLZ}$ to be quite larger than 1 to become a relevant component in our measurement. With such values of $N_{BLZ}$, blazars can provide the bulk of the total UGRB emission (a picture similar to the SFG case).

It is clear that to distinguish between the two interpretations, the model of the shot-noise term is crucial. Physically, this is because we have already observed and cataloged a significant fraction of the closest \g-ray emitting blazars, and thus the possible cross-correlation signal for the unresolved part is generated by a relatively small number of sources, providing a large shot-noise term. 
As mentioned above, the model of the latter depends on predicting the \g-ray luminosity of blazars from their IR luminosity. If the relation obtained in Ref.~\cite{Massaro:2016mgw} extends to the unresolved regime, the conclusion of the \std\ case is likely to hold. On the other hand, a lower \g-ray luminosity for the corresponding IR luminosity would point towards the outcome of the \cpfree\ model. A future dedicated cross-matching analysis of the \Fermi-LAT FL8Y source list with multi-wavelength data could help in clarifying the picture. 

\subsubsection{Dark Matter}
\label{sec:dm}
We now discuss the implications for particle DM. Figs.~\ref{fig:plot2D} and \ref{fig:plot2Dcpfree} show the results for the \std\ and \cpfree\ methods, respectively, for the $b\bar b$ annihilation channel. 
The different samples and methods provide compatible constraints, all excluding annihilation rates higher than (about) the ``thermal'' rate for DM mass of 10 GeV and then increasing with a nearly linear trend for higher masses. The cases of $\tau^+\tau^-$ and $W^+W^-$ final states lead to similar results, whilst constraints for DM annihilating into $\mu^+\mu^-$ are about one order of magnitude weaker, as can be seen in Fig.~\ref{fig:plot2Dch}.

The 1D distributions of the annihilation rate reported in Figs.~\ref{fig:plot2D} and \ref{fig:plot2Dcpfree} show a small peak (in all samples). The peak becomes enhanced and shifts to lower annihilation rates when the low-redshift 2MRS catalog is used, both in the \std\ and \cpfree\ analyses. 
In order to understand if it just a statistical fluctuation or it might be rather a hint for a DM contribution, we deepen the investigation by considering a further subsamples, tailored to the expected behavior of a DM signal.
Ideally, in order to emphasize the DM \g-ray contribution over the astrophysical ones, we need to focus the cross-correlation analysis on a catalog with galaxies at low-$z$ \cite{Fornengo:2013rga}, for galaxies with high-mass and with the lowest possible level of star formation and AGN activity. To these ends, we select 10k galaxies from the \twoMRS catalog (low-$z$), in the corner of high K-luminosity (which corresponds to high mass) and low B-luminosity (which corresponds to low star formation rate). The results are shown in Fig.~\ref{fig:plot1D_lBhK}, where we focus on the $b\bar b$ case and the \std\ analysis, for the sake of brevity.

Interestingly, the peak in the likelihood distribution of the annihilation rate slightly increases in height and moves its position towards lower values of the normalization parameter.
Note that it is the most pronounced peak in the likelihoods of the DM annihilation rate among the different samples. Even though the statistical significance is too low to speculate on the possible presence of a DM contribution, we highlight that in Figs.~\ref{fig:plot2D}, \ref{fig:plot1D_lBhK} and \ref{fig:plot2Dcpfree} the significance tends to increase when considering samples with objects at lower redshift and with higher halo masses, just as expected for a dark matter origin. 
This stimulates to further pursue the particle DM quest exploiting the cross-correlation approach with future data and dedicated studies.

The best-fit for the 2MRS/High-K/Low-B analysis occurs at $M_\chi=37$ GeV and $\sigma v = 4\times\langle \sigma v\rangle_{\rm th}$, therefore in slight excess over the ``natural'' scale. However, as widely discussed in the literature (see, e.g., Ref.~\cite{Regis:2015zka}), the normalization of the DM signal can significantly vary depending on the modeling of the so-called ``boost-factor'' provided by the substructure contribution (because of unknowns in the definition of the minimum halo mass, subhalo mass function and subhalo concentration parameter). Therefore, the normalization of the DM signal can be easily modified by a factor of a few by introducing a substructure modeling different from the one considered here. 

\section{Conclusions}
\label{sec:concl}

In this work, we have made an attempt to characterize the unresolved \g-ray emission of the Local Universe. 
To this aim, we employed {\it Fermi}-LAT skymaps with detected sources being masked and performed the measurement of their angular cross-correlation with the \twoMPZ catalog.
The latter contains about one million of galaxies with a median redshift of 0.07. The cosmological volume probed by \twoMPZ\ powers only about 10\% of the total unresolved \g-ray background. Despite this small fraction, the technique adopted here enables us to study the composition of such emissions.

The null hypothesis, i.e., the absence of correlation between the two datasets, is excluded at a statistical confidence larger than $99.99\%$.

To understand the origin of this correlation, we considered a few different subsamples of the \twoMPZ\ catalog by splitting it into redshift, K-band luminosity (taken as a tracer of the object mass) and B-band luminosity (taken as a tracer of the star formation rate of the object) bins. 
We found misaligned AGNs to be the most likely contributor of the bulk of the signal.
The normalization of this contribution is such that the extrapolation to higher redshift makes mAGN emission compatible with explaining the majority of the UGRB at GeV energies. 
On the other hand, star forming galaxies appear to be a subdominant component in our measurement. Nevertheless, the derived bounds allow them to still be a significant component of the UGRB at higher redshift. 

The energy spectrum of the APS somewhat favors the presence of a blazar-like component at high-energies.
On the other hand, the contribution is rather featureless, being driven by the shot-noise term. In order to fully establish the fraction of their contribution, an improvement in the link between IR and \g-ray luminosity for faint blazars is crucial.

Finally, we evaluated the possible contribution of a particle DM signal.
The $95\%$ C.L. bounds on the DM annihilation rates reach close to the ``thermal'' rate for DM mass of 10 GeV for $b\bar b$, $\tau^+\tau^-$ and $W^+W^-$ annihilation channels (while an order of magnitude weaker bound is found for $\mu^+\mu^-$) and then increasing with a nearly linear trend for higher masses. 
Interestingly, when considering samples where the DM evidence is expected to increase (namely, correlation with objects at low-$z$, with high-mass, and low level of star formation), we see a slightly more pronounced peak in the DM likelihood for the DM contribution. Currently, the statistical significance of this effect is low, and it prevents us from deriving any firm conclusions on the presence of a DM signal. Nevertheless, this result motivates to deepen the investigation of cross-correlations between suitable galaxy catalogs (especially low redshift ones, like 2MRS) and multiwavelength observations, to probe the potential contribution of DM. 

\section*{Acknowledgements}
We would like to thank M. Bilicki, A. Cuoco, F. Massaro and M. Negro for discussions.
This work is supported by the ``Departments of Excellence 2018 - 2022'' Grant awarded by the Italian Ministry of Education, University and Research (MIUR) (L. 232/2016). 
NF is supported by the research grant ``The Anisotropic Dark Universe" Number CSTO161409, funded under the program CSP-UNITO ``Research for the Territory 2016" by Compagnia di Sanpaolo and University of Torino. 
The work of SH is supported by the U.S.~Department of Energy under Award No.~DE-SC0018327. 
MR acknowledges support by the Excellent Young PI Grant: ``The Particle Dark-matter Quest in the Extragalactic Sky'' funded by the University of Torino and Compagnia di San Paolo and by ``Deciphering the high-energy sky via cross correlation'' funded by Accordo Attuativo ASI-INAF n. 2017-14-H.0. 
SA, NF and MR acknowledge support from the project ``Theoretical Astroparticle Physics (TAsP)'' funded by the INFN.

\appendix

\section{Validation and cross-checks}
\label{sec:xck}
In this appendix, we present a series of tests performed in order to validate our analysis. 
\subsection{Theoretical estimation of the error}
It is possible to provide a theoretical estimation of the error $\delta C_l$ associated to the cross-correlation in each multipole bin, assuming gaussian statistics: 
\be
\delta C_l = \sqrt{\frac{(C_l^{(\gamma, {\rm gal})})^2 + C_l^{(\gamma, \gamma)}C_l^{({\rm gal}, {\rm gal})}}{(2l+1)f_{\rm sky}\Delta l}},
\label{eq:theoerr}
\ee
where $f_{\rm sky}$ is the fraction of unmasked sky, $\Delta l$ is the multipole bin size, and $C_l^{(\gamma, {\rm gal})}$, $C_l^{(\gamma, \gamma)}$ and $C_l^{({\rm gal}, {\rm gal})}$ are the cross-correlation, auto-correlation (including noise) of the \g-ray data and auto-correlation (including noise) of the galaxies, respectively. In the top left panel of Fig.~\ref{fig:checkplot} we show the errors on the cross-correlation of the \g-ray data in the energy interval from 1 to 10 GeV with the whole \twoMPZ catalog. We find that the theoretical error of Eq.~\ref{eq:theoerr} is similar to and typically slightly smaller than the one estimated by PolSpice, which we then use throughout our analyses.

\subsection{Foreground dependence}
\label{sec:appfor}

\begin{figure*}[t]
  \centering
  \subfigure{\includegraphics[width=0.3\textwidth]{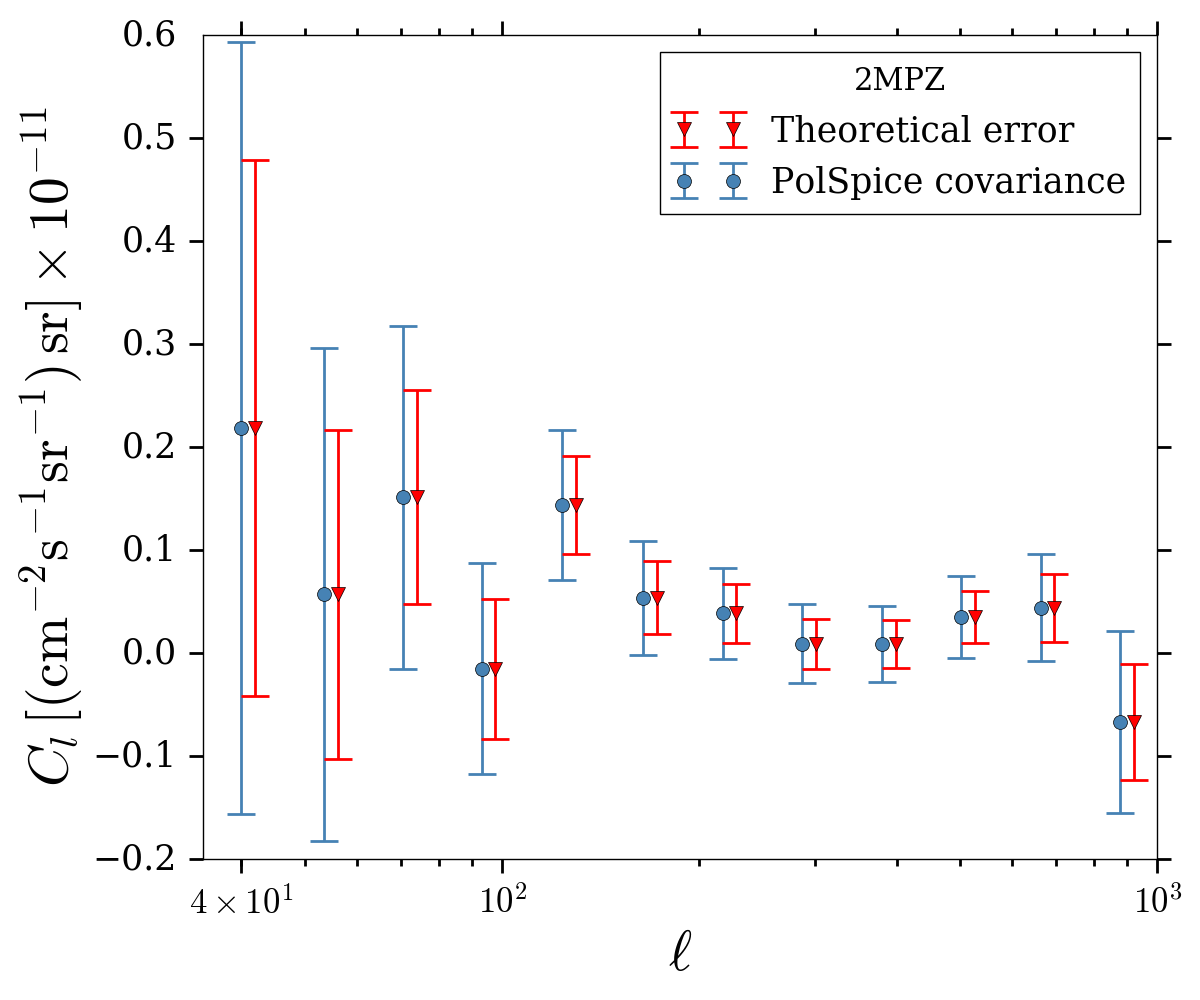}}
  \hspace{0.5cm}
  \subfigure{\includegraphics[width=0.3\textwidth]{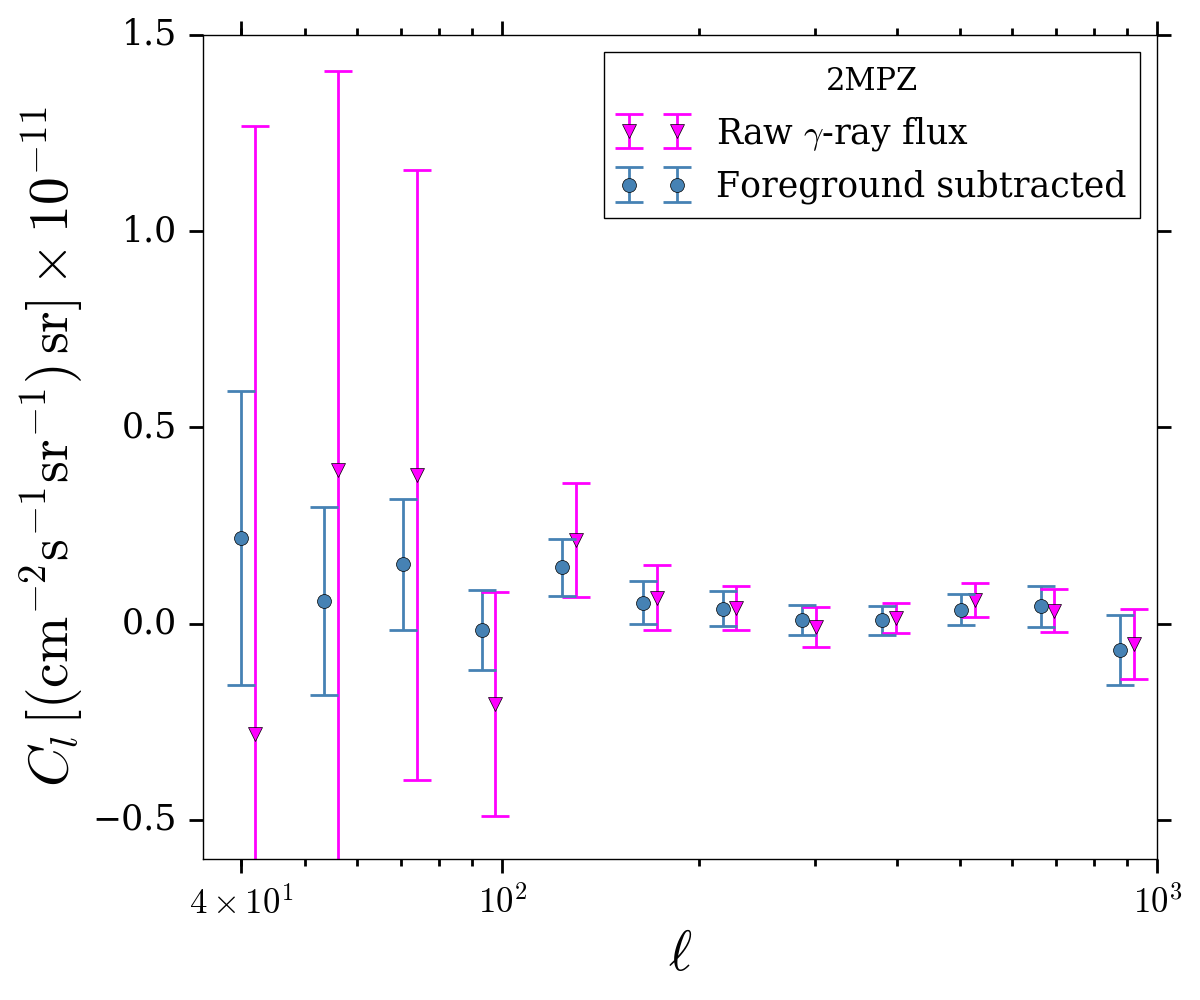}}
  \hspace{0.5cm}
  \subfigure{\includegraphics[width=0.3\textwidth]{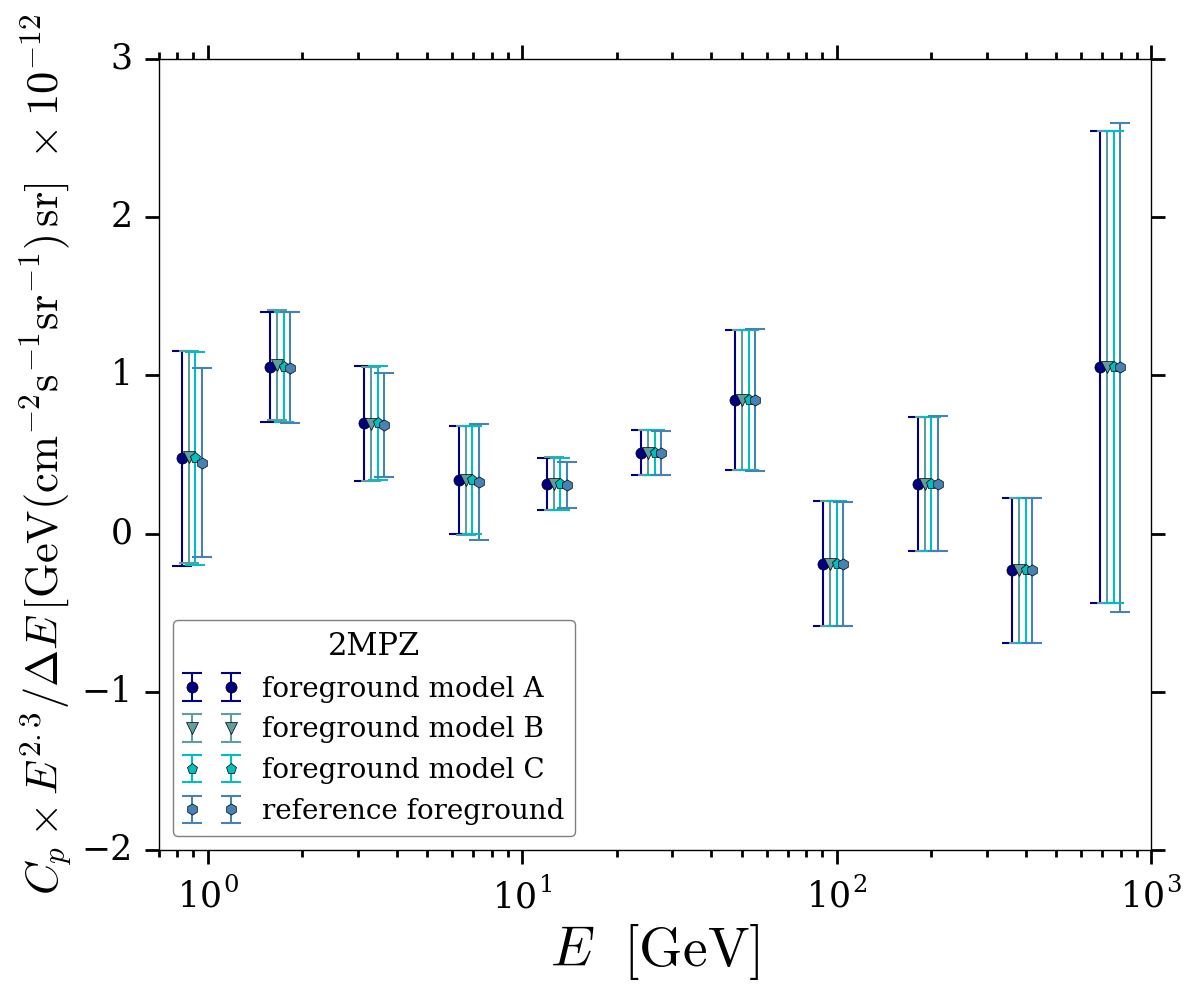}}
  \subfigure{\includegraphics[width=0.32\textwidth]{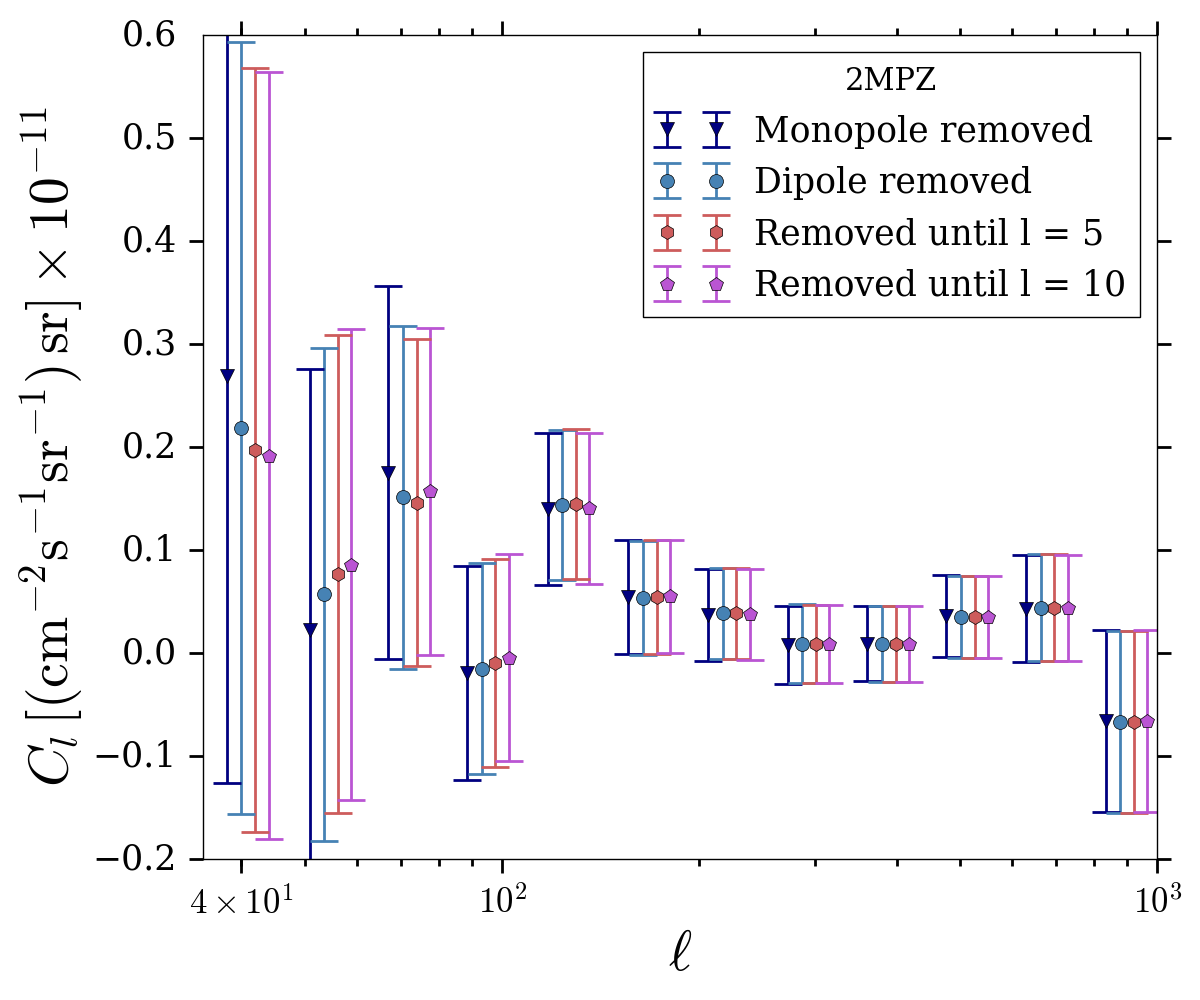}}
  \hspace{2cm}
  \subfigure{\includegraphics[width=0.32\textwidth]{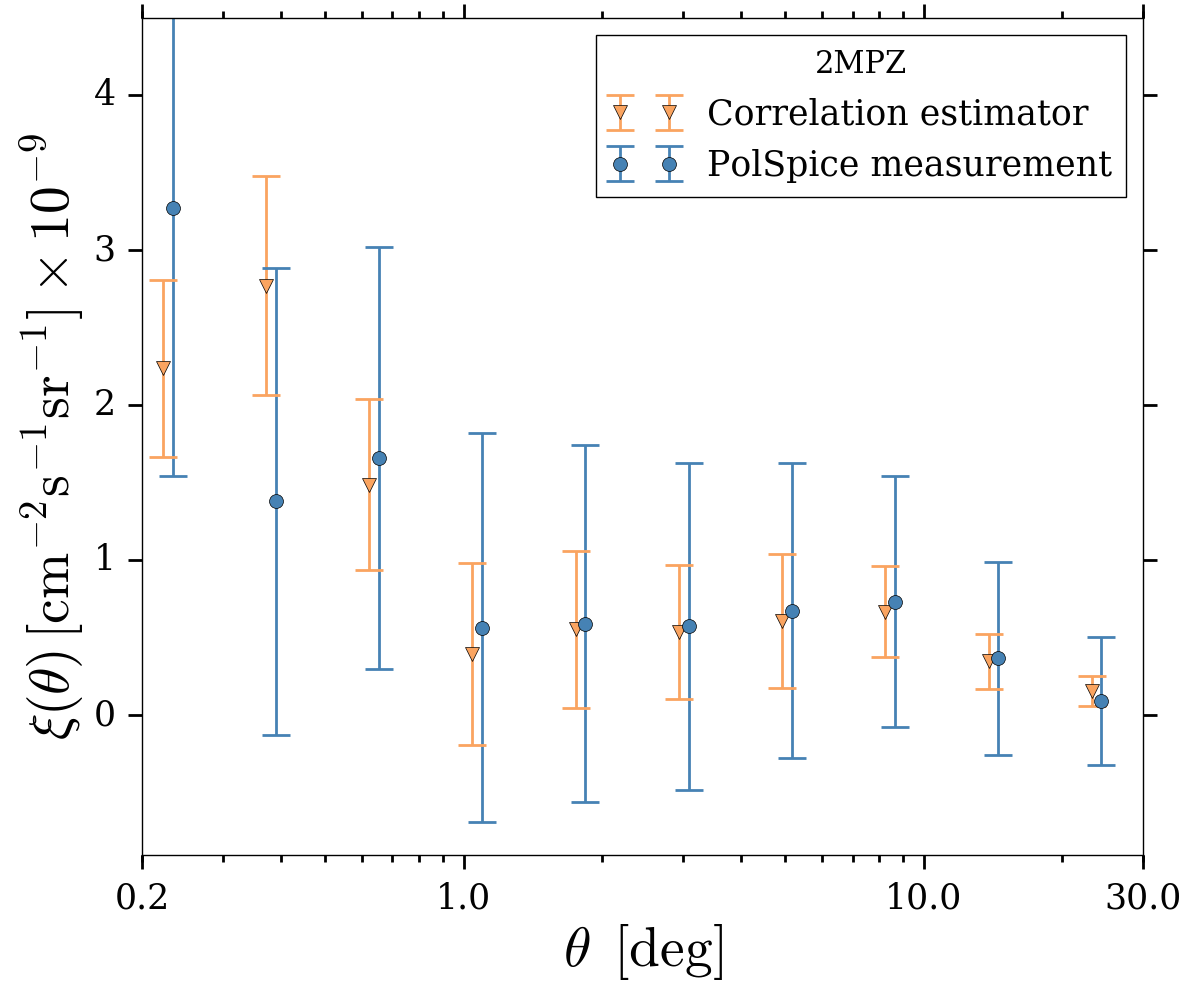}}
  \caption{Tests of stability of our results. 
  Upper left panel: APS and its errors determined with the theoretical Gaussian estimate (red) and the error provided by PolSpice (blue). 
  Upper central panel: APS obtained by using the masked photon maps with (blue) and without (pink) Galactic foreground subtraction.
  Upper right panel: Same as Fig.~\ref{fig:cp} bottom-right (blue-points), but employing different models for the Galactic foreground~\cite{Ackermann:2014usa}.
  Lower left panel: APS obtained by removing the monopole (dark blue), the monopole and the dipole (light blue), the first 5 multipoles (red), and the first 10 multipoles (pink).
  Lower right panel: Angular correlation function measured with the estimator of Eq.~\ref{eq:xi} (yellow) and with PolSpice (blue); the errors for the former are determined by means of a jackknife technique.
All results refer to a $[1,10]$ GeV energy bin (except for upper right panel) and to the full 2MPZ catalog.
  }
  \label{fig:checkplot}
\end{figure*}

In order to assess the independence of our analysis from Galactic \g-ray foreground subtraction, we perform a cross-correlation analysis of the combined energy bins from 1 to 10 GeV (which contains about 60$\%$ of the total photon counts) using \g-ray data that have been cleaned up by the diffuse Galactic emission (as explained in section \ref{sec:fermidata}) and compare those results with a corresponding analysis performed on the same data without foreground removal. The top central panel of Fig.~\ref{fig:checkplot} shows that the APS derived with and without foreground removal are in excellent agreement, which confirms the hypothesis that the cross-correlation of the \g-ray flux with extragalactic tracers of the \g-rays emitters is not affected by the Galactic \g-rays foreground. This suggests that the cross-correlation results are not strongly dependent on the specific foreground model used in foreground removal. 
To further confirm this point, we compute $C_p^k$ as a function of the energy bin as in Fig.~\ref{fig:cp} employing different foreground models in the analysis. In addition to our reference case, we introduce models A, B and C presented in \cite{Ackermann:2014usa}. The differences in our results among the four cases are negligible, as can be seen in Fig.~\ref{fig:checkplot} (top-right) for the example of the full 2MPZ catalog.
Note however from the central panel that the presence of an un-subtracted foreground emission results in a noisier dataset that is reflected in larger uncertainties, especially at lower multipoles. 

\subsection{Lower multipoles removal}

While in a full-sky APS analysis the different multipoles are independent, the presence of masks results in couplings between different multipoles.
PolSpice corrects for this effect, but a residual contamination is still potentially present. Moreover, with the monopole being largely dominant (e.g., for \g-rays the total average intensity is much larger than its fluctuations), even a small residual coupling can bias the measurement at higher multipoles. In our analysis we remove the monopole and the dipole before performing the APS measurement, and we consider the APS only for multipoles larger than 40, as discusses in section \ref{sec:measur}. We nevertheless performed a check to verify that our measurement is not affected by lower multipoles, by comparing the APS results in the multipole window $\ell \geq 40$ when we removed from the map the contribution of multipoles up to $\ell \leq 5$ ($\ell \leq 10$). This is realized with the following procedure: {\it i)} compute the spherical harmonic decomposition coefficients $a_{lm}$ of the maps with the HEALPix routine \code{anafast}; {\it ii)} produce the corresponding skymaps containing only the structure relative to multipoles up to $\ell = 5$ ($\ell = 10$): this is obtained as a constrained realization by feeding the HEALPix routine \code{alm2map} with the $a_{lm}$'s obtained in point {\it i)} only up to $\ell = 5$ ($\ell = 10$); {\it iii)} subtract the maps derived in point {\it ii)} from the original maps. This is an approximate way to subtract lower multipoles, because the effect of the mask is not included in {\it i)} but it is useful to test the impact of possible leakages from low to high multipoles. From these maps we then derive the cross-correlation APS and compare it with the APS determined with only the monopole or monopole and dipole subtracted (the latter is what we do in our baseline analysis). The results are shown in the bottom left panel of Fig.~\ref{fig:checkplot}, which shows that all results are perfectly compatible with each other and  therefore there is no leakage of power from lower multipoles to the multipole window of interest.

\subsection{Correlation in real space}
In order to test the robustness of our measurement, we also compute the cross-correlation function in real space $\xi(\theta)$, which can then be transformed to the APS with the usual relation:
\be
\xi(\theta) = \sum \limits_{l} \frac{(2l+1)}{4 \pi} C_l P_l(\cos\theta),
\ee
where $P_l(\cos\theta)$ are the Legendre polynomials and $\theta$ is the physical angular scale. The correlation function $\xi(\theta)$ is determined by means of the following estimator:
\be
\xi(\theta) = \frac{1}{\sum_{a,b} f_{ab}(\theta)} \sum_{a,b} (n_\gamma - \bar{n}_\gamma) \frac{(n_{\rm gal} - \bar{n}_{\rm gal})}{\bar{n}_{\rm gal}} f_{ab}(\theta),
\label{eq:xi}
\ee
where $(n_\gamma - \bar{n}_\gamma) $ and $(n_{\rm gal} - \bar{n}_{\rm gal})$ represent the fluctuations of the \g-ray intensity flux and of the galaxy number counts in every unmasked $a$-th and $b$-th pixel and the function $f_{ab}(\theta)$ assumes the value 1 when the angular separation of the two pixels is $\theta$ and 0 otherwise.
We compare the correlation function we measure by means of Eq.~\ref{eq:xi} with the corresponding $\xi(\theta)$ provided by PolSpice. The error associated to our estimator is computed with a jackknife re-sampling approach, dividing the sky into 20 distinct patches and estimating the relative covariance. The bottom right panel in Fig.~\ref{fig:checkplot} shows the comparison of the two methods: they nicely agree, with the jackknife method possibly underestimating the errors. 

\subsection{Comparison with previous measurement}
\begin{figure}[t]
  \centering
  \includegraphics[width=1.0\columnwidth]{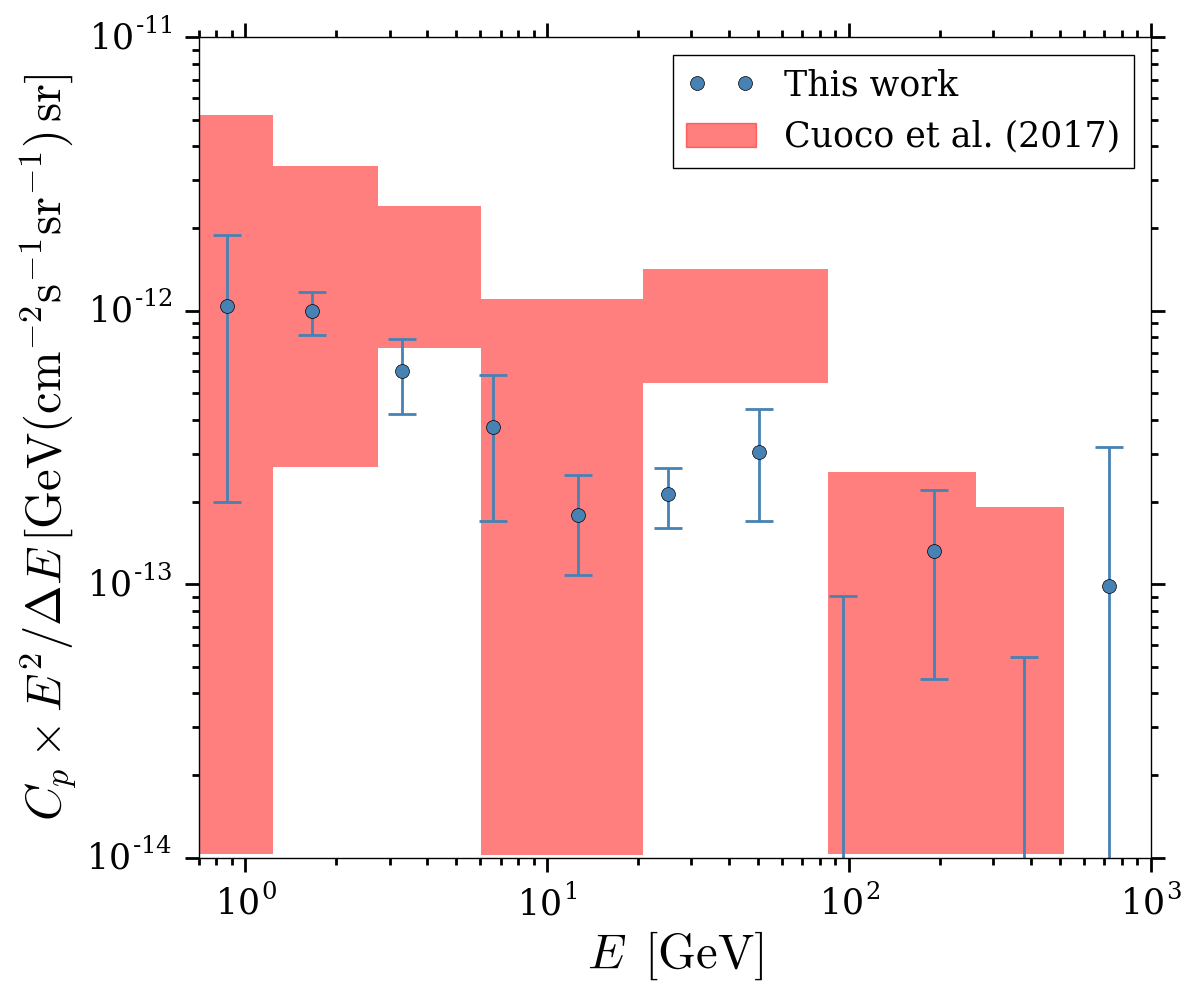}
  \caption{Comparison with previous results for the Poisson noise terms $C_p^k$ as a function of the energy for the full 2MPZ sample. Points refer to our analysis, the shaded regions show the results obtained in the previous analysis of Ref. \cite{Cuoco:2017bpv}.}
  \label{fig:cuoco}
\end{figure}

Finally we compare our measurement with the results obtained in a previous analysis of cross-correlation between \g-rays and the 2MPZ catalog, with a smaller photon statistics \cite{Cuoco:2017bpv}. The comparison is shown in Fig.~\ref{fig:cuoco} and refers to the determination of the Poisson noise terms $C_p^k$ defined in section \ref{sec:ampli} and is performed for the full 2MPZ sample. We see that our results and the results of Ref.~\cite{Cuoco:2017bpv} are in good agreement, and we can appreciate the improvement in the statistical determination of the signal with our new analysis.

\section{Halo occupation distribution of galaxies}
\label{sec:hod}
In this work, we adopt the halo model to describe the clustering of structures. In order to estimate the angular cross-correlation of the unresolved \g-ray sky with samples of galaxies, we need to describe how galaxies populate halos. To this end, we employ the halo occupation distribution (HOD) formalism.

We follow the approach described in Ref.~\cite{Zheng:2004id} (for review on HOD, see also Refs.~\cite{Berlind:2001xk,Cooray:2002dia}), where the HOD is parameterized by distinguishing the contributions of central and satellite galaxies: $N=N_{\rm cen}+N_{\rm sat}$, since different formation histories typically imply different properties for galaxies residing at the centers of halos with respect to satellite galaxies. These can be modeled with the following functional form for the central galaxies:
\be
\langle N_{\rm cen}(M)\rangle = \frac{1}{2}\left[1+{\rm erf}\left(\frac{\log M-\log M_{\rm cut}}{\sigma_{\rm logM}}\right)\right] \label{eq:HOD1}
\ee
and with the following form for the satellite galaxies:
\bea
\langle N_{\rm sat}(M)\rangle &=& \left( \frac{M-M_{\rm cut}}{M_1} \right)^\alpha\; {\rm for}\; M>M_{\rm cut} \nonumber \\
\langle N_{\rm sat}(M)\rangle &=&0\; \;{\rm for}\;\; M\leq M_{\rm cut}\;\;.\label{eq:HOD2}
\eea
With this formalism, we need four parameters for each galaxy population:
$M_{\rm cut}$ denotes the approximate halo mass required to populate the halo with the considered type of galaxy, with the transition from 0 to 1 central galaxy modeled by means of Eq.~(\ref{eq:HOD1}), and set by the width $\sigma_{\rm LogM}$. The satellite occupation is described by a power law (with index $\alpha$ and normalization set by the mass parameter $M_1$.

Eqs.~(\ref{eq:HOD1}) and (\ref{eq:HOD2}) provide the number of galaxies in a halo of mass $M$. Concerning the spatial distribution, we treat central and satellite galaxies separately. The former is taken as a point-source located at the center of the halo (the point-source approximation is expected to break down only for $\ell\gtrsim10^3$).
Satellite galaxies are instead described in an effective way with a spatial distribution following the host-halo profile.
In other words, we express the density field of galaxies with:
\bea
g_g(\bm x - \bm x'|M) = && \langle N_{\rm cen}(M)\rangle\,\delta^3(\bm x - \bm x') + \nonumber \\
&& \langle N_{\rm sat}(M)\rangle\,\rho_h(\bm x - \bm x'|M)/M\, .
\eea
Note that:
\begin{equation}
\int \de^3\bm x\,g_g(\bm x)=\langle N_{\rm cen}(M)\rangle+\langle N_{\rm sat}(M)\rangle=\langle N(M)\rangle\, .
\end{equation}

The value of the four HOD parameters of each sample is derived by fitting the auto-correlation of the specific catalog. We perform the measurement of the auto-correlation by employing the PolSpice tool, in the same way as described in the main text for the cross correlation. The noise term is estimated with $C_N^{gg}=4\,\pi\,f_{\rm sky}/N$, where $N$ is the total number of galaxies outside the mask, and is subtracted from the measurement.

The theoretical prediction for the 3D power spectrum is computed with the halo model approach (and assuming Poisson statistics) as:
\begin{widetext}
\bea
 P_{gg}^{1h}(k,z) &=& \int_{M_{\rm min}}^{M_{\rm max}} dM\ \frac{dn}{dM} \frac{2\langle N_{\rm cen}\,\rangle\,\langle N_{\rm sat}\,\rangle\tilde v_\delta(k|M)+\langle N_{\rm sat}\,\rangle^2\tilde v_\delta(k|M)^2}{\bar n_{g}^2} \label{eq:PSdecLSS1}\\
 P_{gg}^{2h}(k,z) &=& \left[\int_{M_{\rm min}}^{M_{\rm max}} dM\,\frac{dn}{dM} b_h(M) \frac{\langle N_{g}\rangle}{\bar n_{g}} \tilde v_g(k|M) \right]^2\,P^{\rm lin}(k)\;.
\label{eq:PSauto}
\eea
\end{widetext}
The product $\langle N_{g}\rangle\,\tilde v_g(k|M)$ is the Fourier transform of $\langle N_{\rm cen}(M)\rangle\,\delta^3(\bm x )+\langle N_{\rm sat}(M)\rangle\,\rho_h(\bm x|M)/M$. Note that $\langle N_{g}\rangle\,\tilde v_g(k=0|M)=\langle N_{g}\rangle$.
The average number of galaxies at a given redshift is given by $\bar n_{g}(z)=\int dM\,dn/dM\, \langle N_{g}\rangle$.
Note that in Eq.~\ref{eq:HOD1}, we do not include the shot-noise term $\propto\langle N\,\rangle^2$ since it has been subtracted from the data.

\begin{table}[t!]
\begin{center}
\caption{Best fit values of the HOD parameters of Eqs.(\ref{eq:HOD1}) and (\ref{eq:HOD2}) for all the samples considered in this work.}
\label{tab:HOD}
\begin{tabular}{|c|c|c|c|c|}
%\hline
\hline
Catalog  &  $M_{\rm cut}$   & $\sigma_{{\rm Log} M}$ & $\alpha$ &  $M_1$ \\
    &  $[10^{12} M_\odot]$   &  &  &  $[10^{13} M_\odot]$ \\
\hline
   2MPZ (full) & 1.8 & 0.32 & 1.15 & 2.8\\
\hline
   2MRS (full)&   1.6  & 0.22 & 1.0 & 2.0 \\
\hline
2MPZ high-$z$    &   4.6 &  0.32 &  1.2  &  4.4  \\
\hline
2MPZ mid-$z$  & 2.6  & 0.18   & 1.2  &  4.0  \\
\hline
2MPZ low-$z$    &   1.5 &  0.32  & 1.15  & 2.0  \\
\hline
2MPZ high-B   &   2.6 &  0.15  & 1.2  &  3.3 \\
\hline
2MPZ mid-B   &   1.5 &  0.24  & 1.15  &  2.5 \\
\hline
2MPZ low-B   &   0.66 &  0.20  & 1.15  &  1.1 \\
\hline
2MPZ high-K    &   4.6  & 0.30  & 1.2  &  4.4 \\
\hline
2MPZ mid-K    &   1.5  & 0.26  & 1.15  &  2.5 \\
\hline
2MPZ low-K    &   0.50  & 0.28  & 1.15  &  1.1 \\
\hline
2MRS high K - low B  & 2.6  & 0.10   & 1.15    &  2.5  \\
\hline
\end{tabular}
\end{center}
\end{table}

\begin{figure}[t]
\centering
\includegraphics[trim=0 20 0 240,clip, width=0.55\textwidth]
{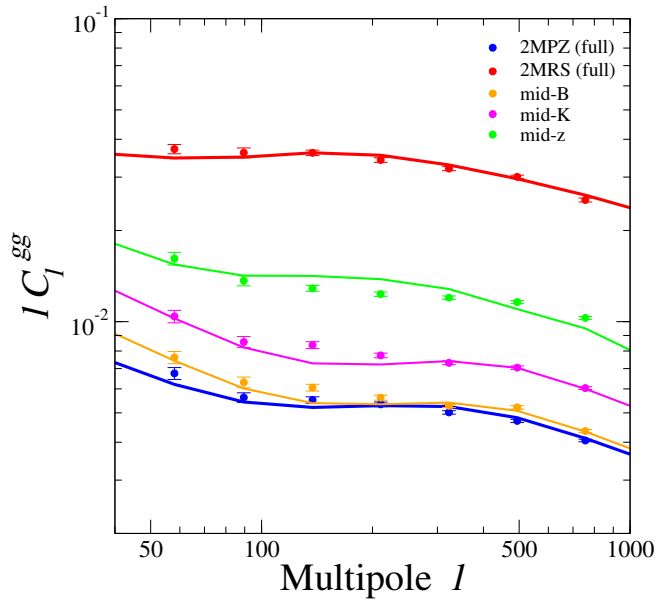}
\caption{Autocorrelation angular power spectrum for the 2MPZ (full, mid-B, mid-K, mid-$z$) and 2MRS catalogs. Points show the measurements, while lines refer to the best-fit model derived as described in the text. } 
\label{fig:autogal}
\end{figure}

The best-fit HOD parameters are reported in Table~\ref{tab:HOD}.
A few examples of the comparison between theoretical model and measured APS are shown in Fig.~\ref{fig:autogal}. It is clear from the plot that the models are strongly constrained by the measurements. Therefore, the uncertainty on the HOD parameters has negligible impact on the cross-correlation observable and can be neglected in our analysis, where we consider only the best-fit values.

\section{Estimate of gamma-ray luminosity from other wavelengths}
\label{sec:cp}

As mentioned in the main text, the Poisson noise term of the cross-correlation signal is given by the average gamma-ray flux of objects in the catalogs. The computation is performed in two steps. First, we derive a relation for the (diffuse) gamma-ray production of all galaxies given some tracer of the star formation rate. Then we add up emissions from blazars and misaligned AGN if the object has been classified as an host of these emitters.

Here we describe how we derive the gamma-ray emission of AGNs and star-forming galaxies starting from a given magnitude in the optical/infrared. Note that such relations suffer from significant uncertainty. If the latter is due just to random scatter around the reported relations, the impact of these uncertainty in our analysis is subdominant. In fact, in order to compute the Poisson noise term, we add up the flux of a very large number of objects. On the other hand, if the adopted relations are biased, this could in principle affect our conclusion. To overcome this issue, we introduce also a model in which the Poisson noise term is not modeled but left free and fitted. 

\subsection{Blazars}
The gamma-ray flux of blazars is computed using the relation between the infrared magnitude at $12\,\mu$m and the energy flux between 0.1 GeV and 100 GeV found in Ref.~\cite{Massaro:2016mgw}. From their Fig.~2, one obtains a $F^E_\gamma=A\, [{\rm W3}]^{-\beta}$ with $A=10^{-14.05\pm0.39} {\rm erg\,cm^{-2}\,s^{-1}}$ and $\beta=4.94\pm0.17$.
We employ the W3 magnitude measured by the WISE survey and provided in the catalog.

\subsection{Misaligned AGNs}
Predictions for the gamma-ray flux of misaligned AGNs are typically derived from their radio emission \cite{DiMauro:2014wha,Hooper:2016gjy}, with the best-fit relation found to be:
$\mathcal{L}_\gamma=10^{-4.044}\,(L_{RC}/{\rm erg\,s})^{1.156}$, where
$\mathcal{L}_\gamma$ is the luminosity between 0.1 GeV and 100 GeV and $L_{RC}$ is
the 5GHz radio core luminosity.
Ref.~\cite{2016MNRAS.462.2631M} shows a correlation between 1.4 GHz luminosity and the $12\,\mu$m luminosity of WISE AGNs (see their Fig.~13).
These two relations allow us to predict the Poisson noise term of mAGNs starting from the W3 magnitude of the 2MPZ catalog.
The predicted average gamma-ray flux agrees well with a more direct estimate we obtained on a smaller sample obtained by cross-matching the 2MPZ sources with the FIRST catalog~\cite{FIRST}, to directly extract radio fluxes (then linked to gamma-ray fluxes using again the relation of Refs.~\cite{DiMauro:2014wha,Hooper:2016gjy}).

\subsection{Star-forming galaxies}
Star formation is expected to trigger gamma-ray production in galaxies.
Indeed, galaxies detected in \g-rays show a tight correlation between the luminosity in the range $(0.1-100)$ GeV and the star formation rate (SFR): $\mathcal{L}_\gamma=(1.3\pm0.3) \times 10^{39}\, ({\rm SFR/M_\odot\,yr})^{1.16\pm0.07}$ erg/s \cite{Ackermann:2012uf}.
In turn, the star formation rate is correlated with the B-band magnitude (see, e.g., Fig.~5 in Ref.~\cite{Bothwell:2009jg}). In Ref.~\cite{Bothwell:2009jg}, they found $L_B=13.7\times 10^9 \,{\rm SFR/M_\odot\,yr} $ with a scatter within one dex.
We estimate the average gamma-ray flux of star forming galaxies starting from the B-band magnitude reported in the 2MPZ catalog and using the above two relations.

\bibliography{references}

\end{document}